\documentclass[preprint]{elsarticle}

\usepackage{amssymb}
\usepackage{amsmath}
\usepackage{mdframed}
\usepackage{hyperref}
\usepackage[dvipsnames]{xcolor}
\usepackage{tablefootnote} % for table footnotes
\usepackage[left=0.6in,right=0.6in,top=1in,bottom=1in]{geometry}
\usepackage[shortlabels]{enumitem}
\usepackage{placeins}
\usepackage{tikz}
\usepackage{capt-of}
\usepackage{multicol}
\let\bs\boldsymbol

\usepackage{subcaption}

\begin{document}

\begin{frontmatter}

\title{On Formulations for Modeling Pressurized Cracks \\ Within Phase-Field Methods for Fracture}

\author[inst1]{Andre Costa}

\affiliation[inst1]{organization={Duke University},%Department and Organization
            addressline={Department of Mechanical Engineering and Materials Science}, 
            city={Durham},
            postcode={27708}, 
            state={NC},
            country={United States}}

\author[inst2]{Tianchen Hu}
\author[inst1]{John E. Dolbow}

\affiliation[inst2]{organization={Argonne National Laboratory},%Department and Organization
            addressline={Applied Materials Division}, 
            city={Lemont},
            postcode={60439}, 
            state={IL},
            country={United States}}

\begin{abstract}

Over the past few decades, the phase-field method for fracture has seen widespread appeal due to the many benefits associated with its ability to regularize a sharp crack geometry.  Along the way, several different models for including the effects of pressure loads on the crack faces have been developed.  This work investigates the performance of these models and compares them to a relatively new formulation for incorporating crack-face pressure loads.  It is shown how the new formulation can be obtained either by modifying the trial space in the traditional variational principle or by postulating a new functional that is dependent on the rates of the primary variables.  The key differences between the new formulation and existing models for pressurized cracks in a phase-field setting are highlighted.  Model-based simulations developed with discretized versions of the new formulation and existing models are then used to illustrate the advantages and differences.  In order to analyze the results, a domain form of the J-integral is developed for diffuse cracks subjected to pressure loads.  Results are presented for a one-dimensional cohesive crack, steady crack growth, and crack nucleation from a pressurized enclosure.  

\end{abstract}


% \begin{highlights}
% \end{highlights}


\end{frontmatter}

\twocolumn{

\section{Introduction}

\label{sec:introduction}

The propagation of pressurized fractures is a physical phenomena of interest or concern in many different fields of engineering. Some examples include hydraulic fracture (fracking) treatments in the oil and gas industry \cite{li2015review, mair2012shale}, pressure vessel rupture \cite{shinmura1997fluid}, fracture in concrete dams \cite{wang2017experimental} and fuel fracture in nuclear reactors \cite{capps2021critical, turnbull2015assessment}. Therefore, predictive simulation tools for this phenomena have been intensively studied in recent years. One of these tools is the phase-field method for fracture \cite{bourdin2000numerical}. Initially developed for traction-free cracks, the method has since been extended to account for pressure loading on the surfaces of cracks, as in \cite{bourdin2012variational, wheeler2014augmented, mikelic2015quasi, peco2017influence, jiang2022phase}.  These various formulations exhibit real differences in terms of their structure and form when it comes to how the pressure loads are incorporated.  
%where only brittle fracture propagation was investigated.
The objective of this work is to examine the impact of the various choices, and to compare them to 
a relatively new formulation for pressurized crack surfaces in a phase-field for fracture context \cite{hu2021variationalthesis}.  
The main contributions of this work are: (a) to show that established formulations for pressure-driven fracture in the phase-field
context have limitations when cohesive processes are involved; (b) to demonstrate that the new formulation, derived from variational principles, can address these limitations and be easily combined with phase-field models of cohesive fracture; and (c) to illustrate the advantages and disadvantages of the various models in terms of accuracy in obtaining various quantities of interest.  

Phase-field methods for fracture regularize sharp crack representations through the use of a scalar phase or damage field whose evolution is governed by minimization principles.  
Such methods first appeared, in different forms, in the works of Bourdin et al. \cite{bourdin2000numerical} and Karma et al. \cite{karma2001phase}. The model introduced in Bourdin et al. \cite{bourdin2000numerical} was obtained by a regularization of the variational formulation of fracture developed in Francfort and Marigo \cite{francfort1998revisiting}, using ideas from Ambrosio and Tortorelli \cite{ambrosio1990approximation}. It has been widely adopted in the mechanics community and extended for use in a variety of fracture mechanics problems,  such as ductile failure \cite{alessi2014gradient, ambati2015phase, miehe2016phase, borden2016phase, hu2021variationalpaper}, hydraulic fracture \cite{wilson2016phase, chukwudozie2019variational, mikelic2015phase1, santillan2018phase, miehe2016phase}, dessication problems \cite{maurini2013crack, heider2020phase, cajuhi2018phase, hu2020frictionless}, dynamic fracture\cite{bourdin2011time, borden2012phase, hofacker2013phase, schluter2014phase, li2016gradient, kamensky2018hyperbolic, moutsanidis2018hyperbolic}, fracture in biomaterials \cite{wu2020fracture, raina2016phase, nagaraja2021phase, gultekin2016phase, gultekin2018numerical} and many more. Some recent reviews can be found in \cite{ambati2015review, wu2020phase, francfort2021variational}.

With regard to the use of the phase-field method for hydraulic fracture problems, one challenge concerns how best to incorporate surface loads that result from pressures on crack faces that are diffuse.  
%One challenge involved in this extension is to properly account for a load applied on the crack faces, as their geometry changes, within a framework where the crack faces are not well-defined due to their diffuse representation. 
One approach is to regularize the resulting surface tractions with an approach that is very similar to how the crack surface energy is regularized.  
%The solution to this is the use of a regularized traction load, using the same ideas that led to the regularization of the crack surface energy. 
Early work along these lines focused on crack surfaces loaded by constant pressures, 
 as in Bourdin et al. \cite{bourdin2012variational} and Wheeler et al.\cite{wheeler2014augmented}. Since these early developments, these models have been used extensively for the study of pressurized fractures, for example in \cite{tanne2022loss, zulian2021large, yoshioka2019comparative, yoshioka2020crack}.
 They were also extended and modified to account for fluid flow inside the fractures and poroelasticity in the surrounding medium \cite{miehe2016phase, mikelic2015phase1, chukwudozie2019variational, wilson2016phase, santillan2018phase, heider2017phase, li2022hydro}. The reader is referred to the recent review by  Heider \cite{heider2021review} for additional works on phase-field methods for hydraulic fracture.  The various models all employ some form of ``indicator function" that assists in the regularization of the surface load itself.  Despite several different indicator functions being proposed, the implication of the particular choice of indicator on the accuracy of the models has yet to be thoroughly examined.  
 
 In this manuscript, a new formulation for the study of pressurized fractures, first proposed in the thesis of Hu~\cite{hu2021variationalthesis} is also examined. In particular, it is studied in combination with a cohesive version of the phase-field for fracture method, which was proposed in the recent works of \cite{lorentz2011convergence, geelen2019phase, wu2017unified}.  This facilitates the study of pressurized fracture in quasi-brittle materials and reduces the sensitivity of the effective strength to the regularization length. To ensure that the cohesive fracture behavior is preserved, the implicit traction-separation law is evaluated for a simple one-dimensional problem and shown to be insensitive to the applied pressure with the new formulation. 
 Fracture initiation and propagation examples are also examined to highlight advantages and limitations of the model. 
 
 As part of the analysis conducted to evaluate the various formulations, the J-integral is used to verify the extent to which mode-I crack propagation occurs when the energy release rate reaches the critical fracture energy.  
 The contour form of the J-integral and its modifications for some common cases of phase-field fracture has been examined by others, see e.g.\ the work of \cite{sicsic2013gradient}, \cite{ballarini2016closed} and \cite{hossain2014effective}.  In the case of pressurized cracks, the contour version of the J-integral is not path independent. Many prior works have focused on developing domain forms of the J-integral for sharp cracks that are domain independent \cite{li1985comparison, shih1986energy}.  In this work, a domain form of the J-integral that is suitable for pressurized phase-field cracks is developed for the first time.  
 
The paper is organized as follows. In Section \ref{sec:model}, a simple model for pressure-induced fracturing is presented and the new phase-field formulation is derived in two different ways.  Section \ref{sec:j_integral} provides the derivation of the domain form of the J-Integral for pressurized phase-field cracks. In Section \ref{sec:fem_implementation}, the discretization scheme using finite elements is presented. Then, in Section \ref{sec:results} some fundamental examples involving crack nucleation and propagation are used to illustrate the performance of the various models and choices of indicator functions. Finally, some concluding remarks and directions for future work are discussed in the last section.
}

\section{Model}
\label{sec:model}

The formulation for treating pressurized cracks in a phase-field setting, first introduced by Hu~\cite{hu2021variationalthesis}, can be derived in two different ways.  In what follows, it is first derived based on energy minimization in quasi-static conditions in subsection \ref{qs_derivation}.
 This illustrates the main difference in the underlying hypothesis for this new model compared to the widely used formulations of \cite{bourdin2012variational} and \cite{mikelic2015quasi}, for example. A second derivation based on a maximum dissipation principle is then provided in subsection \ref{dyn_derivation}. 

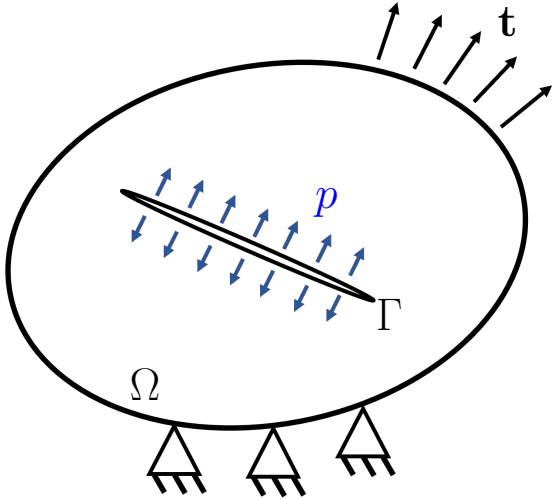
\begin{figure}[ht]
    \centering
    \begin{tikzpicture}
        \node {\pgfimage[interpolate=false,width=.4\textwidth]{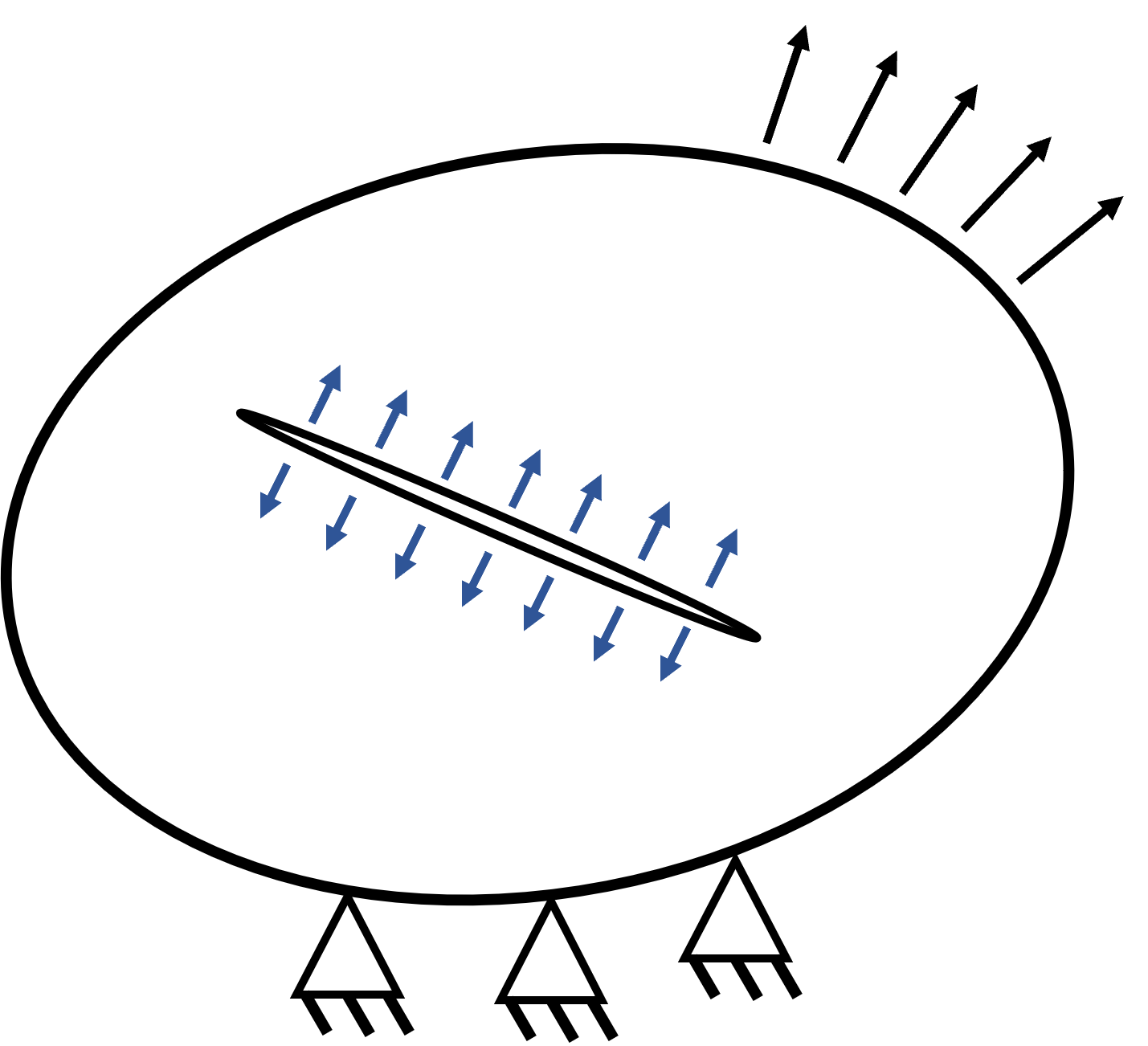}};
        \draw (-0.1\textwidth,-0.1\textwidth) node {\LARGE$\Omega$};
        \draw (0.03\textwidth,0.03\textwidth) node {\LARGE\color{blue}$p$};
        \draw (0.075\textwidth,-0.05\textwidth) node {\LARGE$\Gamma$};
        \draw (0.16\textwidth,0.16\textwidth) node {\LARGE$\textbf{t}$};
    \end{tikzpicture}
    \caption{Generic body containing cracks loaded in pressure.}
    \label{fig:potato}
\end{figure}

 The following assumptions are invoked for both derivations.  
A linear elastic body $\Omega \in \mathbb{R}^n$ ($n = 2$ or $3$), containing cracks denoted by $\Gamma$ is considered (Figure \ref{fig:potato}). The boundary $\partial\Omega$ is partitioned as $\partial\Omega = \partial\Omega_D \cup \partial\Omega_N$, where $\partial\Omega_D$ represents the portion of the boundary where displacements are prescribed and $\partial\Omega_N$ the portion where tractions are applied. Deformations and rotations are assumed to be small, so that a small-strain formulation is appropriate.  For simplicity, body forces are neglected.

\subsection{Quasi-static derivation}\label{qs_derivation}

The quasi-static derivation of the formulation begins by considering the potential energy of a body with cracks which are internally loaded with a pressure $p$.  Crack propagation is associated with a critical fracture energy density, $G_c$.  The total potential energy is given by
\begin{multline}\label{total potential static}
    U(\bs{\epsilon}(\textbf{u})) = \int\limits_{\Omega}\psi_e(\bs\epsilon(\textbf{u}))\ \text{dV} + \int\limits_{\Gamma} G_c \, \text{dA}\\ - \int\limits_{\Gamma} p\textbf{n}\cdot\textbf{u}\ \text{dA} - \int\limits_{\partial \Omega_N} \textbf{t}\cdot\textbf{u} \ \text{dA},
\end{multline}
 in which $\textbf{u}$ are the displacements, $\bs \epsilon(\textbf{u}) = \nabla^s\textbf{u}$ denotes the infinitesimal strain, $\psi_e$ the strain energy density, $\textbf{t}$ the externally applied tractions and $\textbf{n}$ the unit normals of the crack set $\Gamma$ (oriented outwards from $\Omega$).  

In a phase-field for fracture setting, the crack surface $\Gamma$ is regularized with the aid of a scalar phase (or damage) field $d ({\bf{x}}) \in[0,1]$.  In this work, $d = 0$  represents intact material (away from the crack surface) and $d=1$ fully-damaged material (inside the crack).  The damage field is employed in the approximation of the surface integrals in \eqref{total potential static} as volume integrals.  For the energy associated with fracture, several common formulations are encapsulated by the approximation
\begin{equation}
\int\limits_{\Gamma} G_c \, dA \approx  \int\limits_{\Omega} \dfrac{G_c}{c_0\ell}\bigg( \alpha(d) + \ell^2\nabla d \cdot \nabla d\bigg)\ \text{dV}, 
\end{equation}
where $\alpha(d)$ denotes a local dissipation term, $\ell$ is the regularization length, and $c_0$ is a normalization constant given by $c_0 = 4\int_0^1\sqrt{\alpha(s)}ds$.

Such a regularization implies that the distinct crack surface $\Gamma$ is no longer defined.  As such, the second integral on the right of \eqref{total potential static} also needs to be approximated as a volume integral in some manner.  This is effected with the use of an indicator function $I(d)$.  The surface integral involving the pressure is then approximated as
\begin{equation}\label{reg pressure term}
    \int\limits_{\Gamma} p\textbf{n}\cdot{\textbf{u}} \text{dA} \approx \int\limits_{\Omega} p \left( -\frac{\nabla d}{\|\nabla d\|} \right)\cdot{\textbf{u}} \|\nabla I(d)\|\text{dV}.
\end{equation}
 Note that the crack surface normal $\textbf{n}$ is approximated as $-{\nabla d}/{\|\nabla d\|}$, whereas the differential surface element $\text{dA}$ becomes $\|\nabla I\|\text{dV}$.  The indicator function must satisfy $I(0)=0$, $I(1)=1$ and be monotonically increasing. In Bourdin et al.~\cite{bourdin2012variational}, $I(d)=d$ was firstly proposed. Wheeler et al.~\cite{wheeler2014augmented} provide a derivation that avoids an explicit approximation of the normal, such as \eqref{reg pressure term}, but is in fact equivalent to using the indicator function $I(d) = 2d - d^2$. In Peco et al.~\cite{peco2017influence} and Jiang et al.~\cite{jiang2022phase}, $I(d) = d^2$ is used, with the motivation that $I'(0) = 0$ is required to avoid the effects of pressure in undamaged areas.

Combining the approximation in \eqref{reg pressure term} with the traditional phase-field approximation of fracture based on the Ambrosio-Tortorelli functional \cite{bourdin2000numerical} and applying the chain rule, the regularized counterpart of \eqref{total potential static} is given by
\begin{multline}\label{total potential static pf}
    U(\bs{\epsilon},d) = \int\limits_{\Omega}\psi_e(\bs\epsilon,d)\ \text{dV} + \int\limits_{\Omega} p \nabla d\cdot\textbf{u}\ I'(d)\text{dV} \\ + \int\limits_{\Omega}\dfrac{G_c}{c_0\ell}\bigg( \alpha(d) + \ell^2\nabla d \cdot \nabla d\bigg)\ \text{dV} - \int\limits_{\partial \Omega_N} \textbf{t}\cdot\textbf{u} \ \text{dA},
\end{multline}
where the explicit dependence of the strain on the displacements has been dropped.  

Often, the strain energy density is split and part of it is degraded with the damage, i.e. 
\begin{equation}\label{energy split}
    \psi_e(\bs\epsilon(\textbf{u}),d) = g(d)\psi^+_e(\bs\epsilon(\textbf{u})) + \psi_e^-(\bs\epsilon(\textbf{u})),
\end{equation}
where $g(d)$ denotes the degradation function, and $\psi_e^+(\bs\epsilon(\textbf{u}))$ and $\psi_e^-(\bs\epsilon(\textbf{u}))$ denote the ``active" and ``inactive" parts of the energy. The above form encapsulates most of the strain decompositions used in the literature \cite{amor2009regularized},\cite{miehe2010phase} to introduce asymmetry in the fracture behavior in tension and compresssion.

\begin{figure*}[h]
% \centering
\bigskip
\begin{subfigure}{.49\textwidth}
  \centering
  \includegraphics[width=0.7\linewidth]{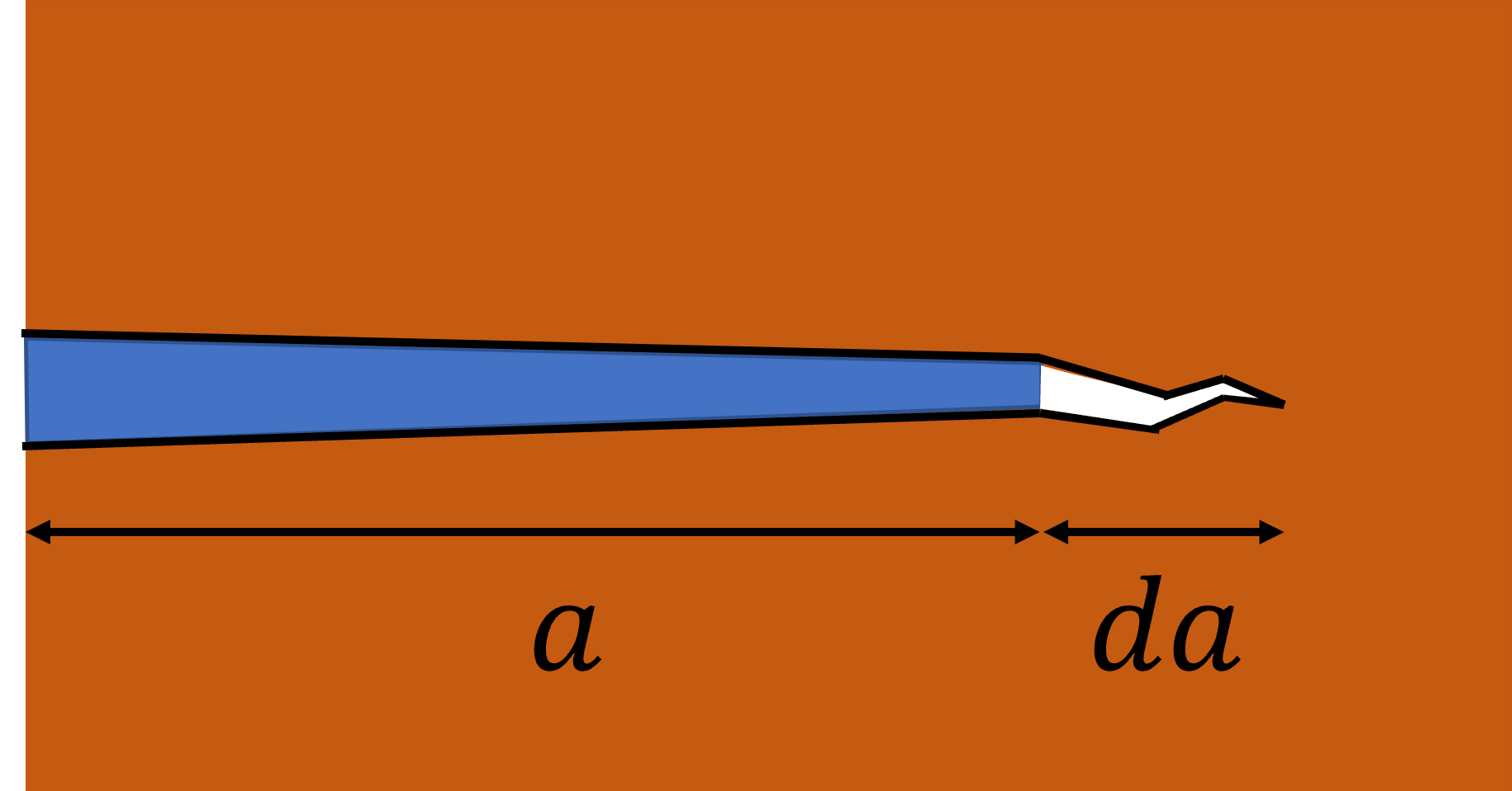}
  \caption{}
  \label{fig:dry_crack}
\end{subfigure}%
\begin{subfigure}{.49\textwidth}
  \centering
  \includegraphics[width=0.7\linewidth]{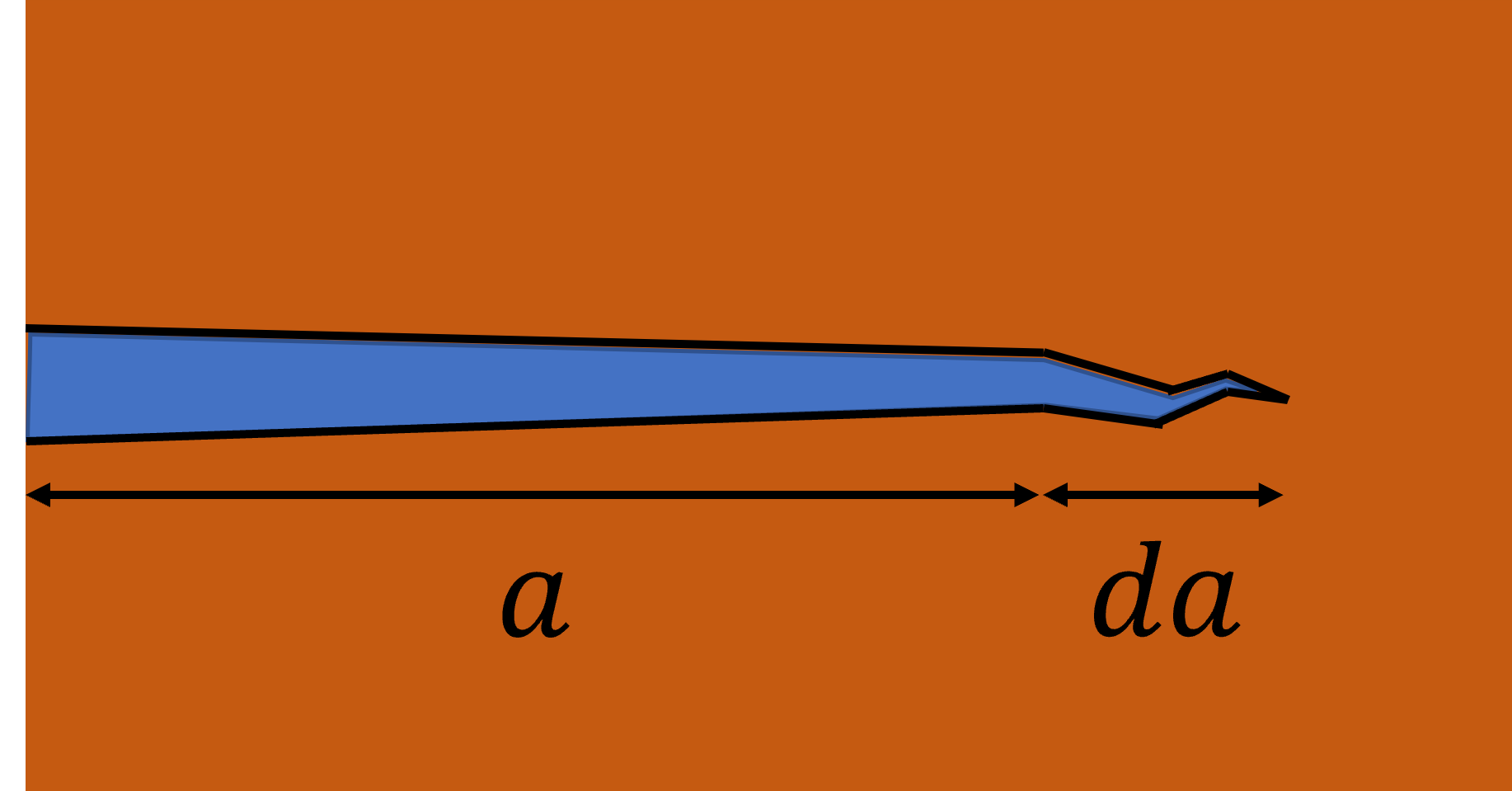}
  \caption{}
  \label{fig:wet_crack}
\end{subfigure}%
  \caption{(a) Unloaded virtual crack. (b) Pressure loaded virtual crack;} 
  \label{fig:wet_vs_dry_crack}
\end{figure*}

Typically, a minimization principle is applied to \eqref{total potential static pf} to extract the governing equations for the displacements $\textbf{u}$ and the damage $d$. According to this principle, a pair ($\textbf{u}$, $d$) is a valid state if and only if all neighboring states ($\textbf{u} + \delta\textbf{u} $, $d+\delta d$) have a greater potential energy. In the case of pressurized cracks, a subtle consideration leads to the formulation proposed herein. Consider the two scenarios indicated in Figure \ref{fig:wet_vs_dry_crack}. In the situation depicted in Figure \ref{fig:wet_crack}, the pressure load (applied in the areas colored in blue), is assumed to 
accompany any crack propagation. Therefore, in an energetic analysis, the virtual crack extension $da$ is assumed  pressurized. By contrast, in Figure \ref{fig:dry_crack}, 
the pressure load is assumed to remain confined to the original crack geometry during propagation.  As a result, the virtual crack extension $da$ is not subject to any surface load.

In terms of the resulting formulation, the difference between the two scenarios shown in Figure \ref{fig:wet_vs_dry_crack} translate into the question of whether or not the damage variation $\delta d$ should enter the pressure work contribution \eqref{reg pressure term}.

For the family of formulations that were developed based on the early work of  \cite{bourdin2012variational} and \cite{wheeler2014augmented}, the scenario depicted in Figure \ref{fig:wet_crack} is assumed as a consequence of including the damage variation $\delta d$ in \eqref{reg pressure term}. The proposed model in this work, by contrast, assumes the case indicated by Figure \ref{fig:dry_crack}.  Although these competing views are expected to give rise to negligible differences in results in the limit as $da \rightarrow 0$, in practice the two formulations do give rise to slightly different sets of governing equations.  As we will demonstrate in the numerical examples in Section~\ref{sec:results}, in practice these differences can translate into fairly significant differences in the results.  

In what follows, the formulation associated with Figure \ref{fig:dry_crack} will be referred to as the Unloaded Virtual Crack formulation, or UVC for short. In the UVC, the variation of the pressure work is simply\footnote{\noindent On the other hand, if the assumption of Figure \ref{fig:wet_crack} is chosen, as in \cite{bourdin2012variational}, two additional terms have to be accounted for, \begin{multline*}\label{wet variation}
    \delta \left( \int\limits_{\Omega} p \nabla d\cdot\textbf{u}\ I'(d)\text{dV} \right) = \int\limits_{\Omega} p \nabla d\cdot\delta\textbf{u}\ I'(d)\text{dV} \\ \underbrace{+ \int\limits_{\Omega} p \nabla \delta d\cdot\textbf{u}\ I'(d)\text{dV} + \int\limits_{\Omega} p \nabla d\cdot\textbf{u}\ \delta d\ I''(d)\text{dV}}_{\text{additional terms}}.
\end{multline*}},

\begin{equation}\label{dry variation}
    \delta \left( \int\limits_{\Omega} p \nabla d\cdot\textbf{u}\ I'(d)\text{dV} \right) = \int\limits_{\Omega} p \nabla d\cdot\delta\textbf{u}\ I'(d)\text{dV}.
\end{equation}

\noindent the variation of the potential energy $\delta U$ can then be written as,

\begin{multline}
    \delta U(\bs{\epsilon},d) = \int\limits_{\Omega}\dfrac{\partial\psi_e}{\partial \bs \epsilon}:\delta\bs\epsilon\ \text{dV} + \int\limits_{\Omega} p \nabla d\cdot\delta\textbf{u}I'(d)\ \text{dV} 
    \\- \int\limits_{\partial \Omega_N} \textbf{t}\cdot\delta\textbf{u} \ \text{dA}
    + \int\limits_{\Omega}g'(d) \psi_e^+(\bs\epsilon)\delta d\ \text{dV}
    \\+ \int\limits_{\Omega}\dfrac{G_c}{c_0\ell}\bigg( \alpha'(d)\delta d + 2\ell^2\nabla d \cdot \nabla \delta d\bigg)\ \text{dV},
\end{multline}
and, with the help of the divergence theorem, 
\begin{multline}
    \delta U(\bs{\epsilon},d) = \int\limits_{\Omega}\biggl(-\nabla \cdot \dfrac{\partial\psi_e}{\partial \bs \epsilon} + pI'(d) \nabla d\biggr)\cdot\delta\textbf{u}\ \text{dV} \\
    + \int\limits_{\partial \Omega_N}\biggl( \dfrac{\partial\psi_e}{\partial \bs \epsilon}\cdot\textbf{n}  -\textbf{t}\biggr)\cdot\delta\textbf{u} \ \text{dA} \\
    + \int\limits_{\Omega}\left(g'(d) \psi_e^+(\bs\epsilon)
    + \dfrac{G_c}{c_0\ell}\alpha'(d) - \nabla \cdot\dfrac{2G_c\ell}{c_0}\nabla d \right)\delta d\ \text{dV}\\ + \int\limits_{\partial\Omega} \dfrac{2G_c\ell}{c_0}(\textbf{n}\cdot\nabla d)  \delta d\ \text{dA}.
\end{multline}

The local minimization principle requires the variation of the potential energy $\delta U$ to be non-negative for any admissible state $\textbf{u},d$. In other words, $\delta U(\bs\epsilon, d) \ge 0$, giving rise to the following equation and boundary condition for $\textbf{u}$, since the variation of the displacement field $\delta \textbf{u}$ is arbitrary:

\begin{equation}\label{disp equation}
    -\nabla \cdot \bs \sigma  + pI'(d) \nabla d = 0 \text{ on }\Omega,
\end{equation}

\begin{equation}\label{disp bcs}
    \bs\sigma \cdot \textbf{n} - \textbf{t} = 0 \text{ on }\partial\Omega_N.
\end{equation}

% \noindent since the variation of the displacement field $\delta u$ is arbitrary, and,
\noindent In the above, $\bs\sigma$ denotes the Cauchy stress, defined as $\bs\sigma = \dfrac{\partial\psi_e}{\partial \bs \epsilon}$. 

For the damage variable, it is assumed that the process is irreversible, such that $\dot{d} \ge 0$.  As such, only positive variations in the damage are admissible, and $\delta U(\bs\epsilon, d) \ge 0$ implies

\begin{equation}\label{damage equation}
    g'(d) \psi_e^+(\bs\epsilon)
    + \dfrac{G_c}{c_0\ell}\alpha'(d) - \nabla \cdot \dfrac{2G_c\ell}{c_0}\nabla d \ge 0 \text{ on }\Omega,
\end{equation}
\begin{equation}\label{damage bcs}
    \dfrac{2G_c\ell}{c_0}\textbf{n}\cdot \nabla d \ge 0 \text{ on }\partial\Omega.
\end{equation}

\noindent Equations \eqref{damage equation} and \eqref{damage bcs} are identical to those of the standard phase-field model for \emph{traction-free cracks}. This is the main difference between the new formulation and existing ones derived from \cite{bourdin2012variational} or \cite{wheeler2014augmented}, for example,  in which the governing equation for the damage field contains additional terms to account for the pressure loads on the virtual cracks. In what follows, we refer to that as the Loaded Virtual Crack formulation, or LVC.  In the boxes below, the governing equations for the UVC are compared to those of the LVC.  

%\begin{minipage}{.48\linewidth}
\begin{mdframed}[
    frametitle={\begin{equation}\label{uvc}\tag{UVC}\text{Unloaded Virtual Crack Formulation}\end{equation}},
    frametitlebackgroundcolor=gray!20,
    backgroundcolor=gray!5,
    linewidth=0pt,
    nobreak=true
  ]
  
\begin{equation}\label{disp equation box}
    -\nabla \cdot \bs\sigma  + pI'(d) \nabla d = 0 \text{ on }\Omega,
\end{equation}

\begin{equation}\label{disp bcs box}
    \bs\sigma\cdot \textbf{n} -\textbf{t} = 0 \text{ on }\partial\Omega_N.
\end{equation}

\begin{equation}\label{damage equation box}
    g'(d) \psi_e^+(\bs\epsilon)
    + \dfrac{G_c}{c_0\ell}\alpha'(d) - \nabla \cdot \dfrac{2G_c\ell}{c_0}\nabla d \ge 0 \text{ on }\Omega,
\end{equation}

\begin{equation}\label{damage bcs box}
    \dfrac{2G_c\ell}{c_0}\textbf{n}\cdot \nabla d \ge 0 \text{ on }\partial\Omega.
\end{equation}
  
\end{mdframed}
%\end{minipage}
\hspace{.03\linewidth}
%\begin{minipage}{.48\linewidth}
\begin{mdframed}[
    frametitle={\begin{equation}\label{lvc}\tag{LVC}\text{Loaded Virtual Crack Formulation}\end{equation}},
    frametitlebackgroundcolor=gray!20,
    backgroundcolor=gray!5,
    linewidth=0pt,
    nobreak=true
  ]
  
\begin{equation}\label{disp equation box2}
    -\nabla \cdot \bs\sigma  + pI'(d) \nabla d = 0 \text{ on }\Omega,
\end{equation}

\begin{equation}\label{disp bcs box2}
    \bs\sigma\cdot \textbf{n} - \textbf{t} = 0 \text{ on }\partial\Omega_N.
\end{equation}

%\smallskip

\begin{multline}\label{wet damage equation box2}
    g'(d) \psi_e^+(\bs\epsilon)
    + \dfrac{G_c}{c_0\ell}\alpha'(d) - \nabla \cdot \dfrac{2G_c\ell}{c_0}\nabla d \\ - \nabla \cdot [p\textbf{u}\ I'(d)] + p\nabla d\cdot \textbf{u}I''(d) \ge 0 \text{ on }\Omega,
\end{multline}
    
\begin{equation}\label{wet damage bcs box2}
    \textbf{n}\cdot \left(\dfrac{2G_c\ell}{c_0}\nabla d+   pI'(d)\textbf{u}\right) \ge 0 \text{ on }\partial\Omega.
\end{equation}
  
\end{mdframed}

It is readily apparent that the governing equations for the displacements are identical in the UVC and the LVC. The main difference is in the absence of the additional pressure-dependent terms in the evolution for the damage field and the accompanying boundary condition.  

In \ref{SIF_equivalence}, an analytical study of the energy release rate of a crack propagating under an arbitrary pressure load $p(x)$ is provided, under the assumptions of the UVC formulation and linear elastic fracture mechanics. The results of the study show that it is possible to recover the classic relationship between the energy release rate and the stress intensity factor with the UVC formulation. This ensures the consistency of the proposed formulation \eqref{uvc} with many theoretical works \cite{detournay2016mechanics, garagash2000tip, detournay2004propagation, garagash2005plane, bunger2005toughness} in the field of hydraulic fracture, where the stress intensity factor is used as the propagation criteria.

\subsection{Derivation using the maximum dissipation principle}\label{dyn_derivation}

In this subsection, an alternative approach to derive the \eqref{uvc} formulation  is presented. It is based on the construction of a total potential functional which depends on the rates of the internal variables $\dot{\bs\epsilon},\dot{d}$ and accounts for  the work of the pressure load as an external dissipation mechanism. This approach is described in more detail in \cite{hu2021variationalpaper} and \cite{hu2021variationalthesis}, where it is used to derive a variationally consistent phase-field model for ductile fracture. 

The total potential is postulated as

\begin{equation}\label{total potential}
    L(\dot{\bs\epsilon},\dot{d}) = \int\limits_{\Omega} \dot{u}(\dot{\bs\epsilon},\dot{d}) \text{dV} - \mathcal{P}^{ext}, 
\end{equation}
 where $u$ is the material internal energy, which relates to the Helmholtz free-energy $\psi$ through $\dot{u} = \dot{\psi} + \dot{T}s$, where $T$ is the temperature and $s$ the entropy. 
In this work, only isothermal processes are considered, therefore, $\dot{u} = \dot{\psi}$. The term $\mathcal{P}^{ext}$ denotes the external power expenditure. If cracks were represented by internal boundaries $\Gamma$ instead of a damage field, one could write,

\begin{equation}\label{power expenditure}
    \mathcal{P}^{ext} = \int\limits_{\partial \Omega \cup \Gamma} \textbf{t}\cdot\dot{\textbf{u}} \text{dA} = \int\limits_{\partial \Omega} \textbf{t}\cdot\dot{\textbf{u}} + \int\limits_{\Gamma} p\textbf{n}\cdot\dot{\textbf{u}} \text{dA}.
\end{equation}

\noindent However, in a regularized setting this integral over $\Gamma$ is once again transformed into a volume integral over $\Omega$, as in \eqref{reg pressure term}, 

\begin{multline}\label{reg power expenditure}
    \int\limits_{\Gamma} p\textbf{n}\cdot\dot{\textbf{u}} \text{dA} \approx \int\limits_{\Omega} p \left( -\frac{\nabla d}{\|\nabla d\|} \right)\cdot\dot{\textbf{u}} \|\nabla I(d)\|\text{dV} \\ = 
    -\int\limits_{\Omega}p\nabla d \cdot\dot{\textbf{u}} I'(d) \text{dV}.
\end{multline}

Recalling the equivalence between the internal energy and the Helmholtz free-energy, the Coleman-Noll procedure can be applied and, in combination with \eqref{power expenditure} and \eqref{reg power expenditure}, leads to the following expression for $L$ as a function of $\psi$: 

\begin{multline}
    L(\dot{\bs\epsilon},\dot{d}) = \\ \int\limits_{\Omega} \left( \dfrac{\partial\psi}{\partial\bs\epsilon}:\dot{\bs\epsilon} + \dfrac{\partial\psi}{\partial d}\dot{d}+\dfrac{\partial\psi}{\partial \nabla d}\cdot {\nabla \dot d} + p\nabla d \cdot\dot{\textbf{u}} I'(d) \right) \text{dV}\\ - \int\limits_{\partial \Omega} \textbf{t}\cdot\dot{\textbf{u}} \text{dA}. 
\end{multline}

The evolution process is postulated to follow the minimizers of this total potential, with the supplemental conditions that damage is an irreversible process and that the displacements $\textbf{u}$ are prescribed over a subset $\partial \Omega_D$ of the boundary. In other words, 

\begin{equation}\label{minimization principle}
    \dot{\bs\epsilon}, \dot{d} = \underset{\dot{\bs\epsilon},\dot{d}}{{\operatorname{argmin}}} \ L(\dot{\bs\epsilon}, \dot{d}) \text{,\ \   subject to } \dot{d} \ge 0 \text{ and } \textbf{u}=\textbf{g} \text{ on }\partial \Omega_D. 
\end{equation}

\noindent Using the Euler-Lagrange equations, the following general evolution equations can then be obtained in terms of the free-energy function $\psi$:
\begin{equation}\label{u_equation}
    \nabla \cdot \dfrac{\partial\psi}{\partial \bs\epsilon} - pI'(d) \nabla d = 0 \text{ on }\Omega,
\end{equation}
\begin{equation}\label{d_equation}
    \nabla \cdot \dfrac{\partial\psi}{\partial \nabla d} - \dfrac{\partial\psi}{\partial d} \ge 0 \text{ on }\Omega,
\end{equation}
with the boundary conditions
\begin{equation}\label{u_bc_helmholtz}
    \dfrac{\partial\psi}{\partial \bs\epsilon}\cdot\textbf{n} - \textbf{t} = 0 \text{ on }\partial\Omega\setminus\partial\Omega_D
\end{equation}
\begin{equation}\label{d_bc_helmholtz}
    \textbf{n} \cdot \dfrac{\partial\psi}{\partial \nabla d} \ge 0 \text{ on }\partial\Omega.
\end{equation}

\noindent To be consistent with the derivation in subsection \ref{qs_derivation}, the Helmholtz free-energy is postulated as,

\begin{equation}\label{helmholtz free-energy postulate}
    \psi(\bs{\epsilon},d) = \psi_e(\bs{\epsilon},d) + \dfrac{G_c}{c_0\ell}\bigg( \alpha(d) + \ell^2\nabla d \cdot \nabla d\bigg),
\end{equation}

\noindent following the regularization based on the Ambrosio-Tortorelli functional. In this case, the general equations \eqref{u_equation}-\eqref{d_bc_helmholtz} take the form
\begin{equation}
\label{eq:UVC-eq2}
    -\nabla \cdot \bs\sigma  + pI'(d) \nabla d = 0 \text{ on }\Omega,
\end{equation}
\begin{equation}
    g'(d) \psi_e^+(\bs\epsilon)
    + \dfrac{G_c}{c_0\ell}\alpha'(d) - \nabla \cdot \dfrac{2G_c\ell}{c_0}\nabla d \ge 0 \text{ on }\Omega,
\end{equation}
 with the boundary conditions
\begin{equation}
    \bs\sigma\cdot \textbf{n}-\textbf{t} = 0 \text{ on }\partial\Omega_N,
\end{equation}
\begin{equation}
\label{eq:UVC-bc2}
    \textbf{n}\cdot \nabla d \ge 0 \text{ on }\partial\Omega.
\end{equation}
By inspection, \eqref{eq:UVC-eq2}-\eqref{eq:UVC-bc2} are identical to \eqref{disp equation}-\eqref{damage bcs}.

\subsection{Constitutive choices of the phase-field formulation}

In the previous subsection, the proposed model for pressurized cracks was developed for a general phase-field regularization of the variational approach to fracture \cite{francfort1998revisiting}, with a free-energy of the form
\begin{equation}\label{PF general form}
    \psi(\bs{\epsilon},d) = \underbrace{g(d)\psi_e^+(\bs{\epsilon}) + \psi_e^-(\bs{\epsilon})}_{\psi_e} + \underbrace{\frac{G_c}{c_0\ell}\bigg( \alpha(d) + \ell^2\nabla d \cdot \nabla d\bigg)}_{\psi_f}.
\end{equation}

\noindent In what follows, the constitutive choices used in the example problems provided in Section \ref{sec:results} are described. 

\subsubsection{Elastic energy and decomposition}

First, in terms of the solid bulk response, an elastic energy of the type \eqref{energy split} is assumed. When the material is undamaged, it reduces to a purely linear elastic energy, that is,

\begin{equation}
    \psi_e(\bs\epsilon(\textbf{u}),0) = \psi_e^+(\bs\epsilon(\textbf{u}),0)+\psi_e^-(\bs\epsilon(\textbf{u}),0) = \dfrac{1}{2}\bs\epsilon(\textbf{u}) : \mathbb{C} : \bs\epsilon(\textbf{u}),
\end{equation}

\noindent where $\mathbb{C}$ is the elasticity tensor.

When damage is present, a decomposition of the energy is often assumed. In many cases, when the applied load to a fracturing body is predominately tensile, the ``no-split" case given by,

\begin{equation}
    \psi_e^-(\bs\epsilon(\textbf{u}),d)=0 \rightarrow \psi_e(\bs\epsilon(\textbf{u}),d) = \dfrac{1}{2}g(d)\bs\epsilon(\textbf{u}) : \mathbb{C} : \bs\epsilon(\textbf{u}),
\end{equation}

\noindent is capable of correctly predicting the material response, while leading to a simpler set of governing equations. However, in a wide-range of scenarios, compressive forces are present, and an energy split is needed to prevent crack formation in zones of high compression, as well as to allow for transmission of compressive forces across fractured faces. 

In Section \ref{sec:results}, one of the example problems will employ the spectral split of Miehe et al. \cite{miehe2010phase},  given by

\begin{align}
    &\psi_e^+(\bs\epsilon(\textbf{u}),d) = \dfrac{1}{2}\lambda\left<\text{Tr }\bs\epsilon\right>_+^2+\mu\bs\epsilon^+:\bs\epsilon^+ \text{, and } \\ &\psi_e^-(\bs\epsilon(\textbf{u}),d) = \dfrac{1}{2}\lambda\left<\text{Tr }\bs\epsilon\right>_-^2+\mu\bs\epsilon^-:\bs\epsilon^-.
\end{align}

Here, $\left< \cdot \right>_+$ and $\left< \cdot \right>_-$ denote the positive and negative parts of a number respectively, while $\bs\epsilon^+$ and $\bs\epsilon^-$ are the positive and negative parts of an additive decomposition of the strain tensor based on the signs of its eigenvalues. A more detailed description, including the derivation of the stiffness matrix in this case, is provided by Jiang et al.\ \cite{jiang2020three}.

\subsubsection{Brittle fracture}

The first and more traditional phase-field model with an energy of the type \eqref{PF general form} was proposed in \cite{bourdin2000numerical}. It was developed to approximate the brittle fracture process of linear elastic materials in the limit of vanishing $\ell$. In its original form, the degradation function 

\begin{equation}\label{quadratic_degradation}
    g(d) = \xi + (1-\xi)(1-d)^2,
\end{equation}

\noindent is used in combination with a quadratic local dissipation $\alpha(d) = d^2$, in what is now called the AT-2 formulation. However, the use of, $\alpha(d) = d$, (widely referred as the AT-1) comes with the advantage of a purely elastic response before the onset of damage and a compactly supported damage field. Therefore, it will be employed in the example in Section \ref{sec:results} where brittle fracture is investigated. The parameter $\xi >0 $ in \eqref{quadratic_degradation} is a small residual stiffness used to avoid a loss of ellipticity in simulations with fully damaged material.  
%In simulations, it is typically taken to be a relatively small value.  

\subsubsection{Cohesive fracture}\label{cohesive_frac}

The phase-field model for cohesive fracture was first proposed by Lorentz et al \cite{lorentz2011convergence, lorentz2011gradient}. In this model, the use of a quasi-quadratic degradation function, given by
\begin{equation}\label{cohesive_degradation}
    g(d) = \xi + (1-\xi)\dfrac{(1-d)^2}{(1-d)^2+md(1+pd)},
\end{equation}
is combined with a linear local dissipation function $\alpha(d) = d$.  The parameter $m$ is defined as $m = \dfrac{G_c}{c_0\ell\psi_c}$, where $\psi_c$ is the nucleation energy, below which no damage is expected to form.  The parameter $p$ is a shape parameter that can be used to adjust the traction-separation response.  In this work, $p=1$ is used.  

\section{A J-Integral for pressurized cracks in a phase-field setting}\label{sec:j_integral}

This Section presents a modified J-integral, capable of retrieving $G$ in the case of pressurized cracks in a phase-field for fracture setting. The resulting integral is then re-cast into a domain-independent form that is more amenable to finite-element calculations.  

A common form of the J-integral, derived for phase-field fracture and applicable to traction-free cracks is given by \cite{sicsic2013gradient,ballarini2016closed} 
\begin{equation}
    J = \textbf{r} \cdot \int\limits_{\zeta} \biggl( \psi(\bs\epsilon, d)\mathbb{I} - \nabla \textbf{u}^T\bs \sigma-\nabla d \otimes \bs\omega \biggr) \textbf{n}\text{ds}, 
\end{equation}
where $\psi(\bs\epsilon, d)$ is given by equation \eqref{PF general form}. In the above, the vector $\textbf{r}$ denotes the crack propagation direction, $\zeta$ is a closed path around the crack tip, $\mathbb{I}$ is the second-order identity tensor, $\textbf{n}$ is the normal to the closed path $\zeta$ and $\bs\omega = \partial \psi/\partial \nabla d = (G_c\ell/c_0)\nabla d$. Compared to the original form of the J-integral proposed by Rice~\cite{rice1968mathematical,rice1968path}, this expression contains additional terms to account for the phase-field parameter $d$. 

Importantly,  Sicsic and Marigo \cite{sicsic2013gradient} show that, under certain conditions, the standard form of the J-integral widely employed for sharp cracks, viz.
\begin{equation}
    J = \textbf{r} \cdot \int\limits_{\zeta} \biggl( \psi_e(\bs\epsilon, d)\mathbb{I} - \nabla \textbf{u}^T\bs \sigma\biggr) \textbf{n}\text{ds},
\end{equation}
 can be used in a regularized phase-field setting. These conditions are:

\begin{enumerate}[start=1,label={\bfseries H\arabic*}]
    \item \label{itm:hyp1}: The regularization length is sufficiently small, so that a separation of scales between the solution in the damage band and the outer solution can be achieved;

    \item \label{itm:hyp2}: The path $\zeta$ intersects the crack plane at a ninety-degree angle;

    \item \label{itm:hyp3}: The path $\zeta$ intersects the crack plane sufficiently far from the crack tip, so that the damage field only varies in a direction perpendicular to the crack plane.   
\end{enumerate}

\noindent In what follows, these same conditions are assumed, as they facilitate a simpler derivation of a modified J-Integral capable of retrieving the energy release rate even in the presence of pressure loads on the crack faces. The main result of this section can then be stated in the following way.

\begin{figure*}[ht]
% \centering
\begin{subfigure}{.49\textwidth}
  \centering

  \includegraphics[width=0.8\linewidth]{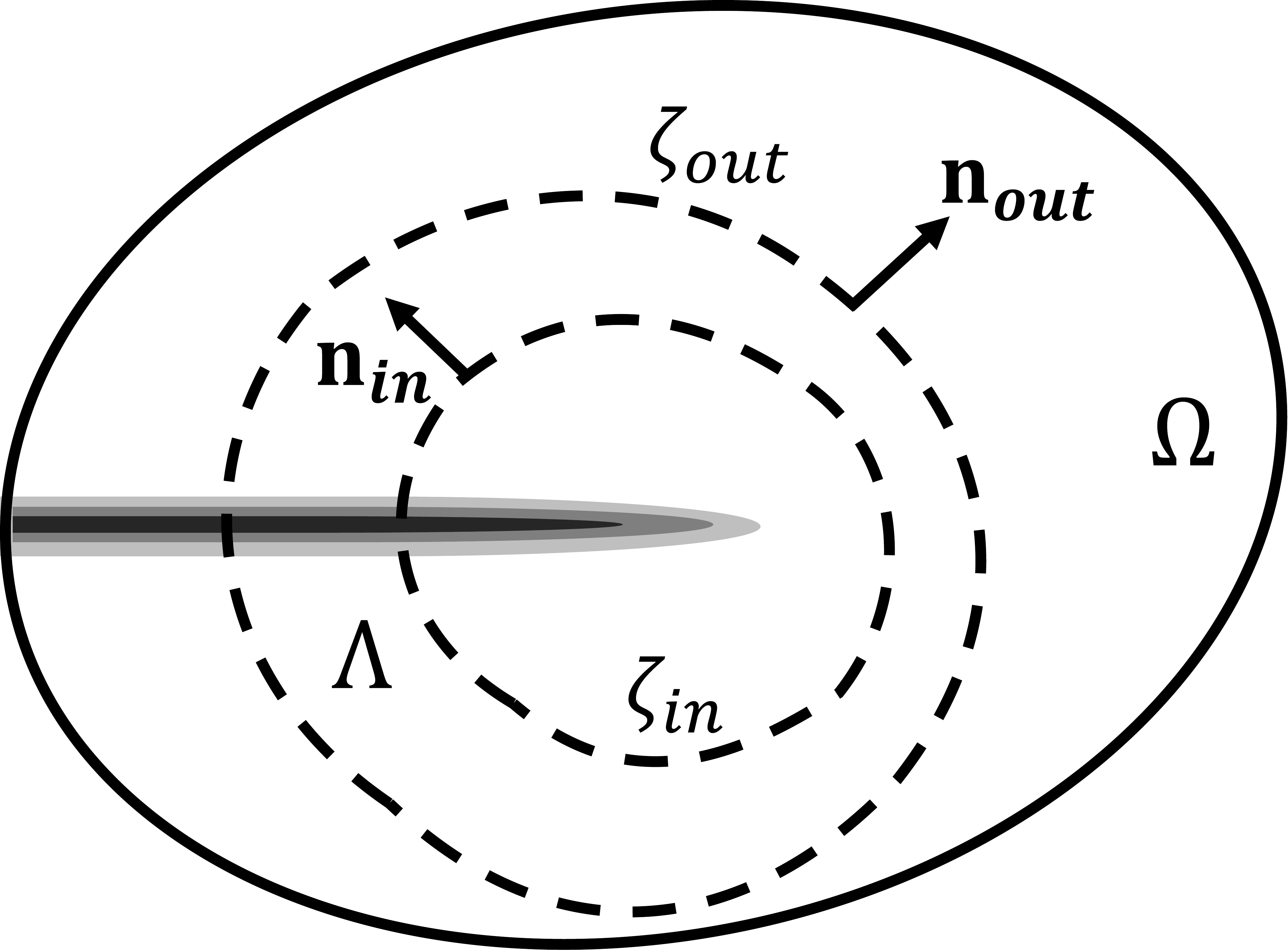}
  \bigskip
  \bigskip
  \caption{}
  \label{fig:j_integral}
\end{subfigure}%
\begin{subfigure}{.49\textwidth}
  \centering
  \includegraphics[width=0.8\linewidth]{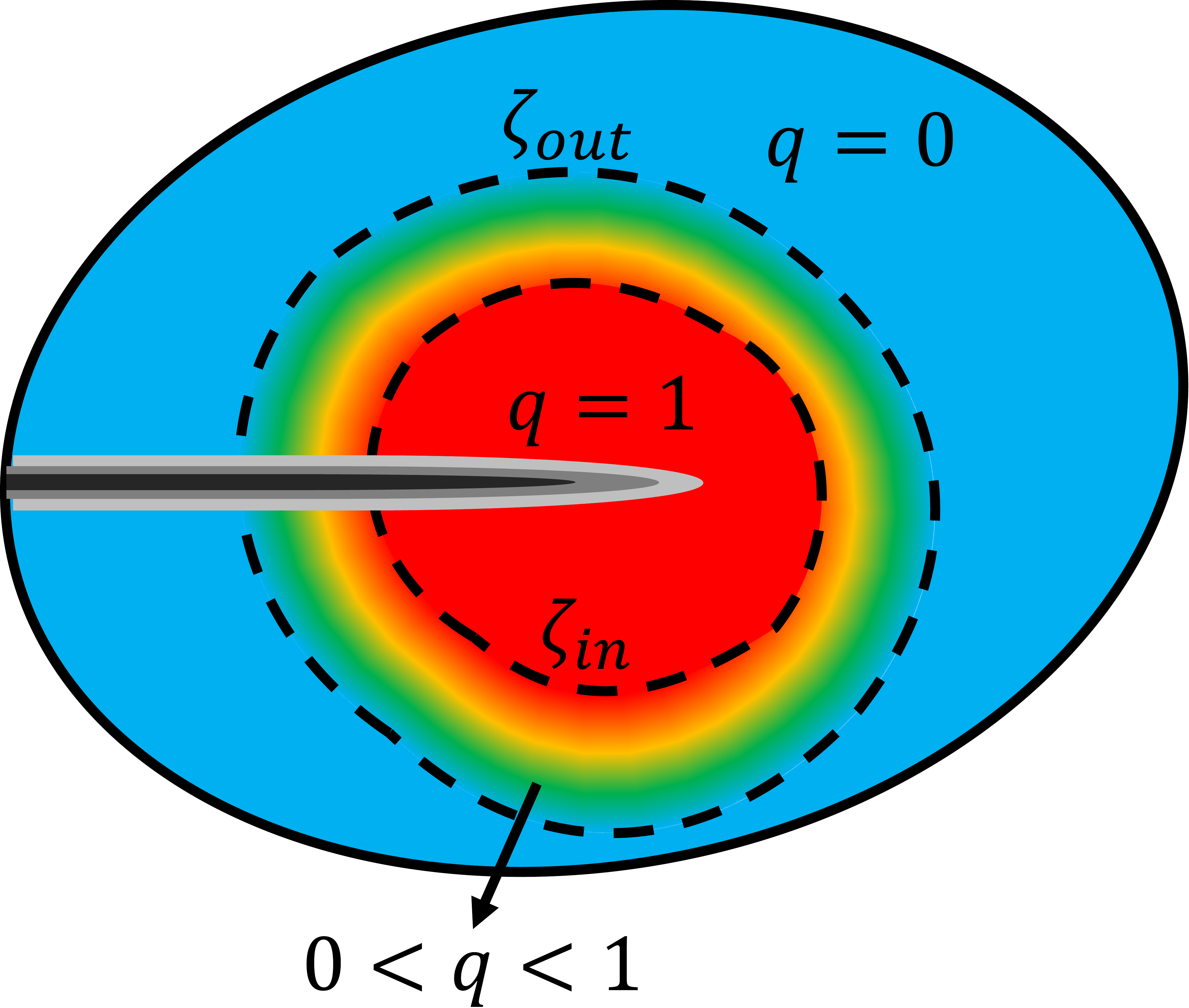}
  \caption{}
  \label{fig:j_integral_q}
\end{subfigure}%
  \caption{(a) The contour paths $\zeta_{in}$ and $\zeta_{out}$ and subdomain $\Lambda$, in the vicinity of a regularized crack.; (b) Color contour plot indicating the assumed variation in the function $q$.  } 
  \label{fig:j_integral_pics}
\end{figure*}

\medskip

\noindent\textbf{Claim}: Consider a domain $\Omega \in \mathbb{R}^2$ with a straight phase-field crack and two closed, non-intersecting paths $\zeta_{in}$ and $\zeta_{out}$ around the crack tip, enclosing an area $\Lambda$ as shown in Figure~\ref{fig:j_integral}. Let $q(x)$ be a sufficently smooth function satisfying $q = 1$ on $\zeta_{in}$ and $q = 0$ on $\zeta_{out}$. Further, assume that $q = 1$ for all points inside $\zeta_{in}$ and $q = 0$ for all points outside $\zeta_{out}$, as shown in Figure \ref{fig:j_integral_q}. Finally, assume that the fracture is loaded by a constant pressure $p$, and that one of the formulations described in Section \ref{sec:model} holds. Then, if conditions \ref{itm:hyp1}, \ref{itm:hyp2} and \ref{itm:hyp3} hold for $\zeta_{in}$ and $\zeta_{out}$, the energy release rate can be approximated by the  integral
\begin{equation}\label{j_integral_theorem}
    J = \textbf{r} \cdot \int\limits_{\Lambda} \biggl( \psi_e(\bs\epsilon, d)\mathbb{I} -p\nabla d\cdot\textbf{u}I'(d)\mathbb{I}  - \nabla \textbf{u}^T\bs \sigma \biggr) \cdot \nabla q\ \text{dA},
\end{equation}

\noindent with an error that vanishes as $\ell \rightarrow 0$. The proof can be established in three steps, as detailed below.

\medskip

\noindent \textbf{Proof:} Define $\textbf{T}_p = \psi_e(\bs\epsilon, d)\mathbb{I} -p\nabla d\cdot\textbf{u}I'(d)\mathbb{I}  - \nabla \textbf{u}^T\bs \sigma$. Using the divergence theorem, one can show that
\begin{equation}
    \int\limits_{\Lambda} \nabla\cdot(q\textbf{T}_p)\text{dA} = \int\limits_{\zeta_{in}\cup\zeta_{out}}q\textbf{T}_p\cdot\textbf{n}\ \text{ds} = 0 - \int\limits_{\zeta_{in}}\textbf{T}_p\cdot\textbf{n}\ \text{ds}, 
\end{equation}
because $q$ vanishes on $\zeta_{out}$ and the normal $\textbf{n}$ to $\zeta_{in}$ points inward to $\Lambda$, as shown in Figure~\ref{fig:j_integral}.  
%sign change in the last term is due to the orientation of the normal $\textbf{n}$. 
Multiplying both sides by the crack direction $\textbf{r}$ and applying the chain rule yields
\begin{equation}\label{first_step}
    -\textbf{r}\cdot\int\limits_{\Lambda} \biggl( \textbf{T}_p\cdot \nabla q + q\nabla \cdot \textbf{T}_p\biggr)\text{dA} =  \textbf{r}\cdot\int\limits_{\zeta_{in}}\textbf{T}_p\cdot\textbf{n}\ \text{ds}, 
\end{equation}
 which completes the first step of the proof. 
 
 The second step consists of showing that the term $R = \textbf{r}\cdot\int\limits_{\Lambda} q\nabla \cdot \textbf{T}_p\ \text{dA}$ is zero. Expanding the expression for $\textbf{T}_p$ and using the chain rule yields
\begin{multline}
    R = \textbf{r}\cdot\int\limits_{\Lambda} q\nabla \cdot \biggl( \psi_e(\bs\epsilon, d)\mathbb{I} -p\nabla d\cdot\textbf{u}I'(d)\mathbb{I}  - \nabla \textbf{u}^T\bs\sigma \biggr)\ \text{dA} = \\
    \int\limits_{\Lambda} q\biggl( \dfrac{\partial \psi_e(\bs\epsilon, d)}{\partial\nabla^s\textbf{u}}\cdot(\nabla(\nabla^s\textbf{u})\textbf{r}) + \dfrac{\partial \psi_e(\bs\epsilon, d)}{\partial d}\nabla d\cdot \textbf{r} 
    \\ - (\nabla \cdot \bs\sigma)\cdot(\nabla\textbf{u}\textbf{r}) - \bs\sigma(\nabla (\nabla\textbf{u})\textbf{r})-p\nabla d\cdot(\nabla \textbf{u}\textbf{r})I'(d)\\ - p\textbf{u}\nabla(I'(d)\nabla d)\textbf{r} \biggr)\ \text{dA}.
\end{multline}

 Re-arranging some terms, one can write,
 
\begin{multline}
    R = \int\limits_{\Lambda} q\biggl[\biggl( \dfrac{\partial \psi_e(\bs\epsilon, d)}{\partial\nabla^s\textbf{u}}-\bs\sigma\biggr) \cdot(\nabla(\nabla^s\textbf{u})\textbf{r}) 
    \\-\biggl( \nabla \cdot \bs\sigma - pI'(d) \nabla d \biggr)\cdot (\nabla\textbf{u}\textbf{r})
    + \dfrac{\partial \psi_e(\bs\epsilon, d)}{\partial d}\nabla d\cdot \textbf{r} 
    \\- p\textbf{u}\nabla(I'(d)\nabla d)\textbf{r} \biggr]\ \text{dA}.
\end{multline}

\noindent For any elastic material, the definition of stress implies $\bs\sigma = \dfrac{\partial \psi_e(\bs\epsilon, d)}{\partial\nabla^s\textbf{u}}$, and due to equation \eqref{u_equation}, $\nabla \cdot \bs\sigma - p \nabla d = 0$, so, the expression above reduces to
\begin{equation}
    R = \int\limits_{\Lambda} q\biggl(\dfrac{\partial \psi_e(\bs\epsilon, d)}{\partial d}\nabla d\cdot \textbf{r} 
    - p\textbf{u}\nabla(I'(d)\nabla d)\textbf{r} \biggr)\ \text{dA}.
\end{equation}
Assuming a separation of scales, the domain $\Lambda$ can be separated into two regions: (i) $\Lambda_{band}$, which consists of the intersection between $\Lambda$ and the support of the damage field representing the crack and (ii) $\Lambda_{outer}$, which denotes the remainder of $\Lambda$, outside of the damage band.  In the asymptotic limit as $\ell\rightarrow 0$, the material in $\Lambda_{outer}$ behaves as purely elastic. Within $\Lambda_{outer}$, one has $d \approx \nabla d \approx 0$, and therefore,  
\begin{equation}
    R_{outer}  = \int\limits_{\Lambda_{outer}} q\biggl(\dfrac{\partial \psi_e(\bs\epsilon, d)}{\partial d}\nabla d\cdot \textbf{r} - p\textbf{u}\nabla(I'(d)\nabla d)\textbf{r} \biggr)\ \text{dA} \approx 0.
\end{equation}
 For the $\Lambda_{band}$ region, by condition \ref{itm:hyp3}, $\nabla d$ is purely perpendicular to the crack direction, so, $\nabla d \cdot \textbf{r} \approx 0$. Therefore,
\begin{equation}
    R_{band} = \int\limits_{\Lambda_{band}} q\biggl(\dfrac{\partial \psi_e(\bs\epsilon, d)}{\partial d}\nabla d\cdot \textbf{r} 
    - p\textbf{u}\nabla(I'(d)\nabla d)\textbf{r} \biggr)\ \text{dA} \approx 0.
\end{equation}

\noindent Since $\Lambda = \Lambda_{band} \cup \Lambda_{outer}$, we must have 

\begin{equation}\label{step2}
    R = R_{outer}+R_{band} \approx 0.
\end{equation}

\noindent This completes the second step.

The final step of the proof begins by invoking the separation of scales to decompose the contour integral in \eqref{first_step} via
\begin{equation}\label{step3}
    \textbf{r}\cdot\int\limits_{\zeta_{in}}\textbf{T}_p\cdot\textbf{n}\ \text{ds} = \textbf{r}\cdot\biggl(\int\limits_{\zeta^{band}_{in}}\textbf{T}_p\cdot\textbf{n}\ \text{ds} + \int\limits_{\zeta^{outer}_{in}}\textbf{T}_p\cdot\textbf{n}\ \text{ds} \biggr).
\end{equation}
On the $\zeta^{outer}_{in}$ portion of the path, damage effects can be neglected and the integral simplifies to the standard (sharp) J-Integral.  In the case of a uniformly pressurized crack \cite{karlsson1978jintegral}, this gives

\begin{equation}\label{path_outer}
    \int\limits_{\zeta^{outer}_{in}}\textbf{T}_p\cdot\textbf{n}\ \text{ds} = G - pw,
\end{equation}
where $w$ denotes the crack aperture at the intersection of the crack and the contour $\zeta_{in}$. 

The other portion of the integral can be re-written as
\begin{multline}
    \textbf{r}\cdot \int\limits_{\zeta^{band}_{in}}\textbf{T}_p\cdot\textbf{n}\ \text{ds} = \\ \textbf{r}\cdot\int\limits_{-B}^{B}(\psi_e\mathbb{I}-\nabla \textbf{u}^T\bs\sigma)\cdot\textbf{n}dx - \textbf{r}\cdot\int\limits_{-B}^{B}p(\nabla d\cdot \textbf{u}I'(d))\cdot\textbf{n}dx,
\end{multline}

\noindent where $B$ is the half-length of the damage band and condition \ref{itm:hyp2} is used to transform the integral over $\zeta^{band}_{in}$ to a simple real integral from $-B$ to $B$. Here, both $\textbf{r}$ and $\textbf{n}$ are unit vectors that point in opposite directions, and therefore, $\textbf{r}\cdot\textbf{n}=-1$, so,

\begin{equation}
    \textbf{r}\cdot \int\limits_{\zeta^{band}_{in}}\textbf{T}_p\cdot\textbf{n}\ \text{ds} = \int\limits_{-B}^{B}(\psi_e\mathbb{I}-\nabla \textbf{u}^T\bs\sigma)dx + p\int\limits_{-B}^{B}(\nabla d\cdot \textbf{u}I'(d))dx.
\end{equation}
Following \cite{bourdin2012variational}, the second integral on the right approaches the crack aperture $w$ as the regularization length decreases, while the first integrand on the right is bounded \cite{sicsic2013gradient}, and therefore this term is $O(B)$, so,

\begin{equation}\label{path_band}
    \textbf{r}\cdot \int\limits_{\zeta^{band}_{in}}\textbf{T}_p\cdot\textbf{n}\ \text{ds} = O(B) + pw = O(\ell) + pw,
\end{equation}

\noindent since the damage band half-length $B$ scales with the regularization length $\ell$. One can now go back to \eqref{step3}, and substitute \eqref{path_outer} and \eqref{path_band} to obtain,

\begin{equation}\label{third_step}
    \textbf{r}\cdot\int\limits_{\zeta_{in}}\textbf{T}_p\cdot\textbf{n}\ \text{ds} = G + pw - pw + O(\ell).    
\end{equation}

\noindent Finally, combining \eqref{first_step}, \eqref{step2} and \eqref{third_step}, one obtains

\begin{equation}\label{end_step}
    -\textbf{r}\cdot\int\limits_{\Lambda}  \textbf{T}_p\cdot \nabla q\ \text{dA} =  G + O(\ell), 
\end{equation}

\noindent which concludes the proof.

\section{Finite Element Implementation}
\label{sec:fem_implementation}

In this Section, the details of the finite element discretization used to obtain approximations to the solution of the proposed model \eqref{uvc} are described. For analogous equations for the model \eqref{lvc}, the reader is referred to \cite{jiang2022phase}.

First, the strong form of the governing equations, derived from the general free-energy \eqref{PF general form} using the KKT\cite{karush1939minima, kuhn1951nonlinear} conditions is presented.

\begin{mdframed}[
  frametitle={Strong form},
  frametitlebackgroundcolor=gray!20,
  backgroundcolor=gray!5,
  linewidth=0pt,
  nobreak=true
  ]
  \vspace{-1em}
  %\small{
  \begin{align}
    &\text{Linear momentum balance: }\nonumber \\ & \nabla \cdot \bs\sigma -p\nabla d + \textbf{b} = 0,                       \hspace{2cm} \forall x \in \Omega, \\ 
                                     & \bs\sigma = \frac{\partial\psi_e}{\partial \bs\epsilon},   \hspace{3.8cm}                 \forall x \in \Omega, \\
                                     & \bs\sigma\cdot \textbf{n} = \textbf{t},                             \hspace{4.2cm}              \partial_t\Omega,  
                                     \\
                                     & \textbf{u} = \textbf{u}_g,                                                  \hspace{4.4cm}     \partial_u\Omega ,                \\
    &\text{Fracture evolution: }\nonumber \\      
    & \dot d\left(\nabla\cdot\frac{2G_c\ell}{c_0}\nabla d - \frac{G_c}{c_0\ell}\alpha'(d) - g'(d)\psi_e^+ (\bs\epsilon)\right) = 0,                                        \forall x \in \Omega,           \\
    & \nabla \cdot \frac{2G_c\ell}{c_0}\nabla d - \frac{G_c}{c_0\ell}\alpha'(d) - g'(d)\psi_e^+ (\bs\epsilon) \le 0,                                         \forall x \in \Omega,  \label{damage ineq}         \\
                                     & \dot d \ge  0,                                     \hspace{4.4cm}     \forall x \in \Omega, \\
                                     & \nabla d\cdot\textbf{n}_0 = 0,                      \hspace{3.8cm}       \partial\Omega, \\
                                         & d(0,\textbf{x}) = d_0,                         \hspace{3.9cm}           \Omega.                                  
    \end{align}
    %}
\end{mdframed}

For the derivation of an equivalent weak form, trial spaces for $\textbf{u}$ and $d$ are first defined.  Although the derivation is confined to quasi-static loadings, the spaces are indexed by a discrete load step parameter $t$.   The trial spaces are given by

  \begin{align}
    \boldsymbol{\mathcal{U}}_t & = \{ \textbf{u} \in \mathcal{H}^1(\Omega)^d \mid \textbf{u} = \overline{\textbf{u}}_t \text{ on } \partial_u\Omega \}, \\
    \mathcal{D}_t      & = \{ d \in \mathcal{H}^1(\Omega) \mid d_{t-1}(x) \le d_t(x) \le 1,\ \forall x \in \Omega \}, 
  \end{align}

\noindent and the accompanying weighting spaces $\boldsymbol{\mathcal{V}}$ and $\mathcal{C}$ are

  \begin{align}
    \boldsymbol{\mathcal{V}} & = \{ \textbf{w} \in \mathcal{H}^1(\Omega)^d \mid \textbf{w} = \boldsymbol{0} \text{ on } \partial_u\Omega \}, \\
    \mathcal{C}      & = \{ c \in \mathcal{H}^1(\Omega) \mid c(x) \ge 0,\ \forall x \in \Omega \}.                                                     
  \end{align}

The condition of monotonicity in the space $\mathcal{D}_t$ is used to prevent damage healing and is the weak enforcement of the condition $\dot d \ge 0$, in a time discrete setting. Denoting the inner product in $\mathcal{H}^1(\Omega)$ and $\mathcal{H}^1(\Omega)^d$ by $\left( \cdot, \cdot \right)$ and it's restriction in the boundary by $\left<\cdot,\cdot\right>$, the weak form of the problem can be written as

\begin{mdframed}[
    frametitle={Weak form},
    frametitlebackgroundcolor=gray!20,
    backgroundcolor=gray!5,
    linewidth=0pt,
    nobreak=true
  ]
  Find $\textbf{u} \in \boldsymbol{\mathcal{U}}_t$ and $d \in \mathcal{D}_t$, such that $\forall \textbf{w} \in \boldsymbol{\mathcal{V}}$ and $\forall c \in \mathcal{C}$,

  \begin{multline}
            \left( \nabla \textbf{w}, \bs\sigma \right) - \left( \textbf{w}, p\nabla d \right) - \left( \textbf{w}, \textbf{b} \right) - \left< \textbf{w}, \textbf{t} \right>_{\partial_t\Omega} = 0,
  \end{multline}

    \begin{multline}\label{weak damage equation}
            \frac{2\ell}{c_0}\left( \nabla c, G_c\nabla d \right) + \frac{1}{c_0\ell}\left( c, G_c\alpha'(d) \right) \\ + \left( c, g'(d)\psi_e^+ (\bs\epsilon(\textbf{u})) \right) = 0,
  \end{multline}
   
\noindent with the initial damage condition,
 
    \begin{align}
      \left( c, d(0,\textbf{x}) - d_0 \right)                & = 0.
    \end{align}
 
\end{mdframed}

Observe that \eqref{weak damage equation} is an equality rather than an inequality, such as \eqref{damage ineq}.  This reflects a view ahead, toward discretization, where in the present work  the irreversibility constraint is enforced with an active-set strategy.  The active set strategy effectively partitions the domain into active (where $\dot{d}=0$) and inactive (where $\dot{d}>0$) parts. Only the inactive part requires a discretization of the damage condition \eqref{damage ineq}, where it is indeed treated as an equality. A detailed description of this constrained optimization algorithm is given by Heister et al. in \cite{heister2015primal}, and some additional details pertinent to phase-field for fracture discretizations can be found in Hu et al. \cite{hu2020phase}.

Finally, these function spaces can be discretized over a finite element mesh, that give rise to the discrete function spaces $\boldsymbol{\mathcal{U}}^h_t \subset \boldsymbol{\mathcal{U}}_t$, $\boldsymbol{\mathcal{V}}^h \subset \boldsymbol{\mathcal{V}}$, $\mathcal{D}^h_t \subset \mathcal{D}_t$, $\mathcal{C}^h \subset \mathcal{C}$. These are then used to construct the discrete form of the problem using the Galerkin method:

\begin{mdframed}[
    frametitle={Spatially discretized form},
    frametitlebackgroundcolor=gray!20,
    backgroundcolor=gray!5,
    linewidth=0pt,
    nobreak=true
  ]
  Find $\textbf{u}^h \in \boldsymbol{\mathcal{U}}^h_t$ and $d^h \in \mathcal{D}^h_t$, such that $\forall \textbf{w}^h \in \boldsymbol{\mathcal{V}}^h$ and $\forall q^h \in \mathcal{C}^h$,

   \begin{multline}\label{eq: semidiscrete momentum balance}
     \left( \nabla \textbf{w}^h, \bs\sigma^h \right) - \left( \textbf{w}^h, p\nabla d^h \right) - \left( \textbf{w}^h, \textbf{b} \right) - \left< \textbf{w}^h, \textbf{t} \right>_{\partial_t\Omega} = 0,
   \end{multline} 

   \begin{multline}\label{eq: semidiscrete fracture evolution}
     \frac{2\ell}{c_0}\left( \nabla c^h, G_c\nabla d^h \right) + \frac{1}{c_0\ell}\left( c^h, G_c\alpha'(d^h) \right) \\ + \left( c^h, g'(d^h)\psi_e^+ (\bs\epsilon(\textbf{u}^h)) \right) = 0,
   \end{multline}    
  
\noindent with the initial damage condition,

\begin{align}
  \left( c^h, d^h(0,\textbf{x}) - d_0 \right)                & = 0.
\end{align}

\end{mdframed}

In this work, bilinear finite elements are used to approximate the damage and displacement fields. 

The coupling between the two discrete equations \ref{eq: semidiscrete momentum balance} and \ref{eq: semidiscrete fracture evolution} is handled by an alternating minimization scheme. A detailed description of this scheme is given in \cite{hu2020phase}.   This solution scheme is implemented using RACCOON \cite{raccoon}, a  parallel finite element code specializing in phase-field fracture problems. RACCOON is built upon the MOOSE framework \cite{gaston2009moose, permann2020moose} developed and maintained by  Idaho National Laboratory.

\section{Results}
\label{sec:results}

We now present results for a set of problems that highlight the advantages, as well as some limitations, of the various models for pressurized cracks in a phase-field for fracture setting. In the first problem, the cohesive fracture of an uniaxial specimen in a pressurized environment is analyzed. We then consider the problem of crack nucleation from a pressurized hole in a media subjected to far-field, biaxial compression.  

Finally, a crack propagation example is studied to verify that in the limit of a vanishing regularization length Griffith-like behavior is recovered with the new model.  In all cases, plane-strain conditions are assumed to hold.  

In the course of explaining the results obtained with the cohesive phase-field model, it will be useful to characterize the effective cohesive strength $\sigma_c$ of the material.  To that end we will rely on the following relationship between the cohesive strength and the nucleation energy:
\begin{equation}
  \label{eq:sigmacrit-from-psicrit}
   \sigma_c = \sqrt{ \dfrac{2E \psi_c }{(1-\nu^2)} },
\end{equation}
where $E$ denotes Young's modulus and $\nu$ Poisson's ratio.  This equation results from the analysis of a one-dimensional system subjected to uniaxial loading \cite{geelen2019phase}, and should be viewed as an approximation to the cohesive strength in more general loading conditions.  

\subsection{Uniaxial bar under traction in a pressurized environment}

We consider the fracture behavior of a cohesive material with pressure loading on the crack faces.  The example is intended to examine the extent to which the pressure loading can artificially influence the apparent traction-separation law on the crack surface.  

\begin{figure}[h]
    \centering
    \includegraphics[width=\linewidth]{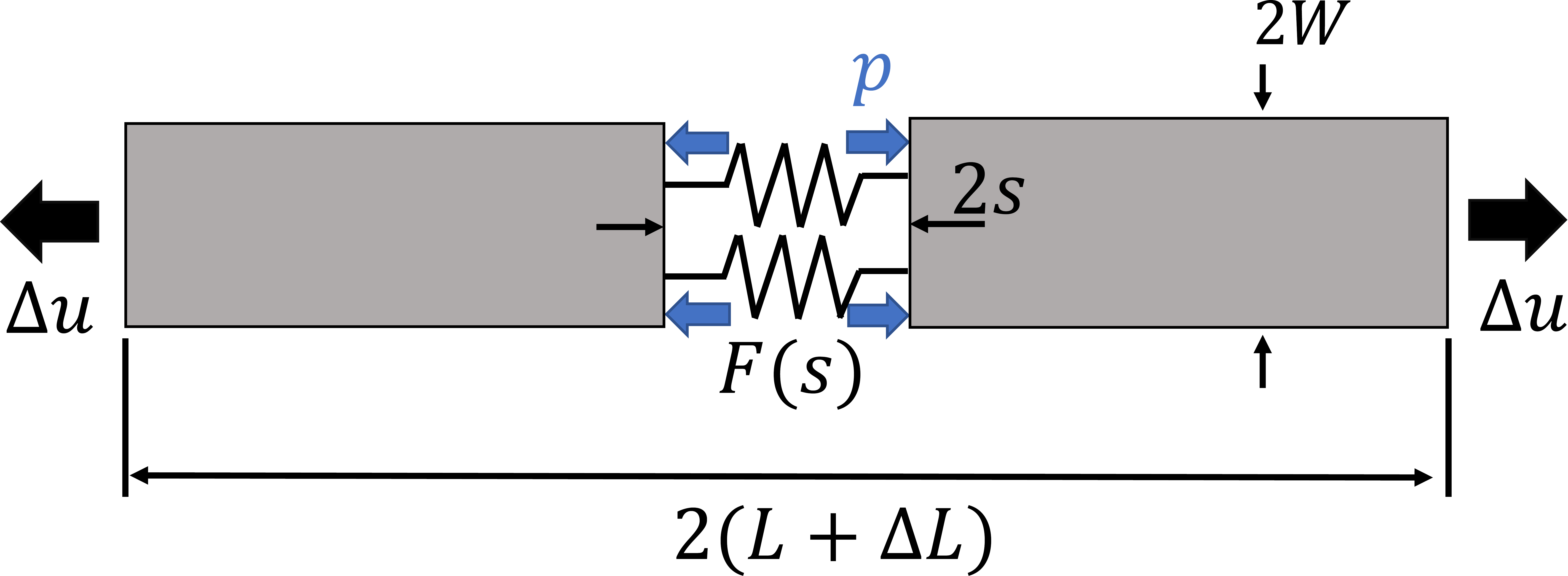}
    \caption{Uniaxial cohesive bar.}
    \label{fig:1d_problem_schematic}
\end{figure}

\begin{table}[h]
\centering
\caption{Material properties for uniaxial bar}
\begin{tabular}[t]{lcc}
\hline
&Value &Unit \\
\hline
Young's modulus ($\text{E}$)&4.0$\times10^5$&MPa\\
Poisson's ratio ($\nu$)&0.2&--\\
 Nucleation energy  ($\psi_c$)&5.6$\times10^{-5}$&$\text{mJ mm}^{-3}$\\
Critical fracture energy ($G_c$)&0.12&$\text{mJ mm}^{-2}$\\
%Tensile strength ($\sigma_c$)&3.0&$\text{MPa}$\\
Residual stiffness ($\xi$)&1.0$\times10^{-8}$&--\\
\hline
\end{tabular}
\label{material_properties_p1}
\end{table}%

The problem consists of a bar under a displacement controlled load in a pressurized chamber, as shown in Figure~\ref{fig:1d_problem_schematic}. The bar is assumed to be made of a linear elastic material that undergoes cohesive fracture, with a traction-separation law $F(s)$.

The bar has an undeformed length $2L = 400$ mm and width $2W = 2$ mm. The material properties are given in Table \ref{material_properties_p1}. Symmetry boundary conditions are invoked to reduce the computational domain to the top-right quarter of the bar. The applied load is modeled as a displacement boundary condition on the right end of the domain. The mesh consists of rectangular elements of size $h$ along the length direction and size 1 mm in the width direction.
The initial applied displacement increment is $\Delta u = 5\times10^{-4}$ mm. The displacement increment is adaptively refined when convergence is not obtained within a fixed set of iterations.   A more detailed description of the adaptive stepping procedure is provided in \cite{gaston2009moose, permann2020moose, lindsay20222}. 

Damage localization is triggered by introducing an small initial defect ($d = \mathcal{O}(\epsilon)$) on the left side of the domain.  In what follows, results are reported using $\ell = L/20 = 10$ mm and $h = \ell/10 = 1$ mm.  This choice of regularization length and mesh spacing was found to yield spatially-converged results. 
Different values of pressure, ranging from $0$ to $\sigma_c/3$ are considered. 

The problem is simulated using discretized versions of both the \ref{uvc} and \eqref{lvc} formulations.  
 
\begin{figure*}[h]
\centering
\begin{subfigure}{.45\textwidth}
  \centering
  \includegraphics[width=\linewidth]{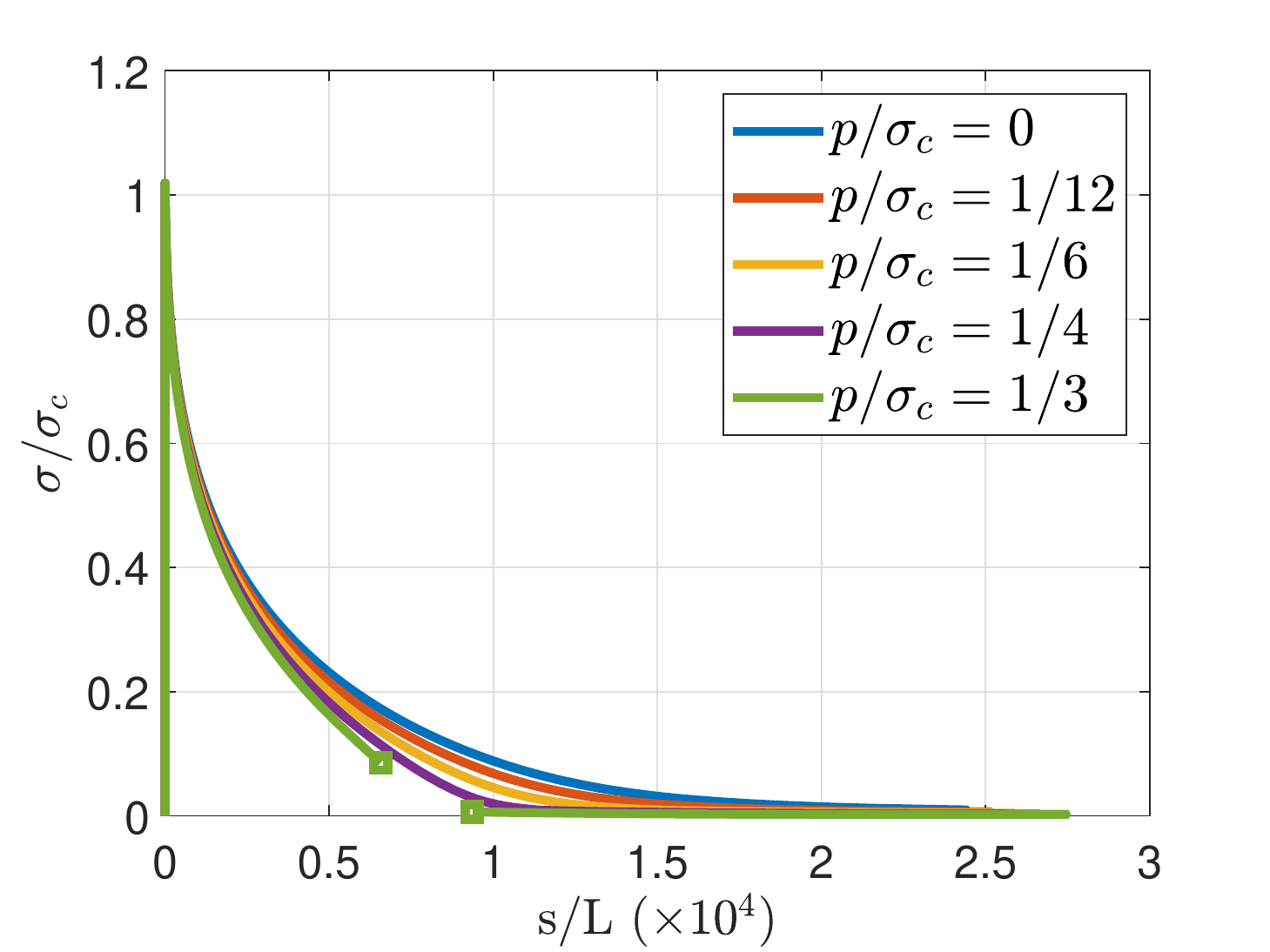}
  \caption{}
  \label{fig:traction_separation_bourdin_d}
\end{subfigure}%
\begin{subfigure}{.45\textwidth}
  \centering
  \includegraphics[width=\linewidth]{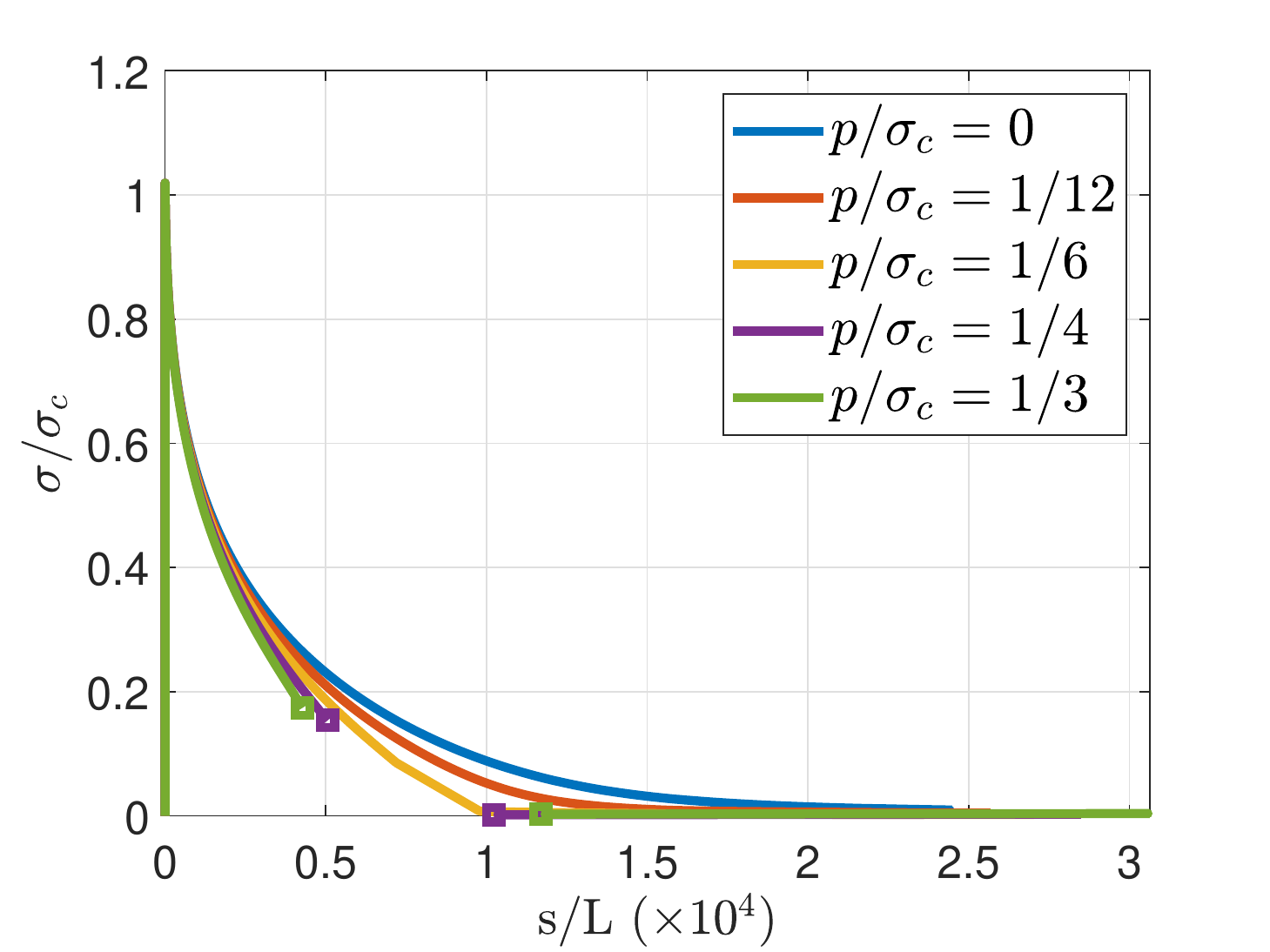}
  \caption{}
  \label{fig:traction_separation_bourdin_d2}
\end{subfigure}%

\bigskip
\begin{subfigure}{.45\textwidth}
  \centering
  \includegraphics[width=\linewidth]{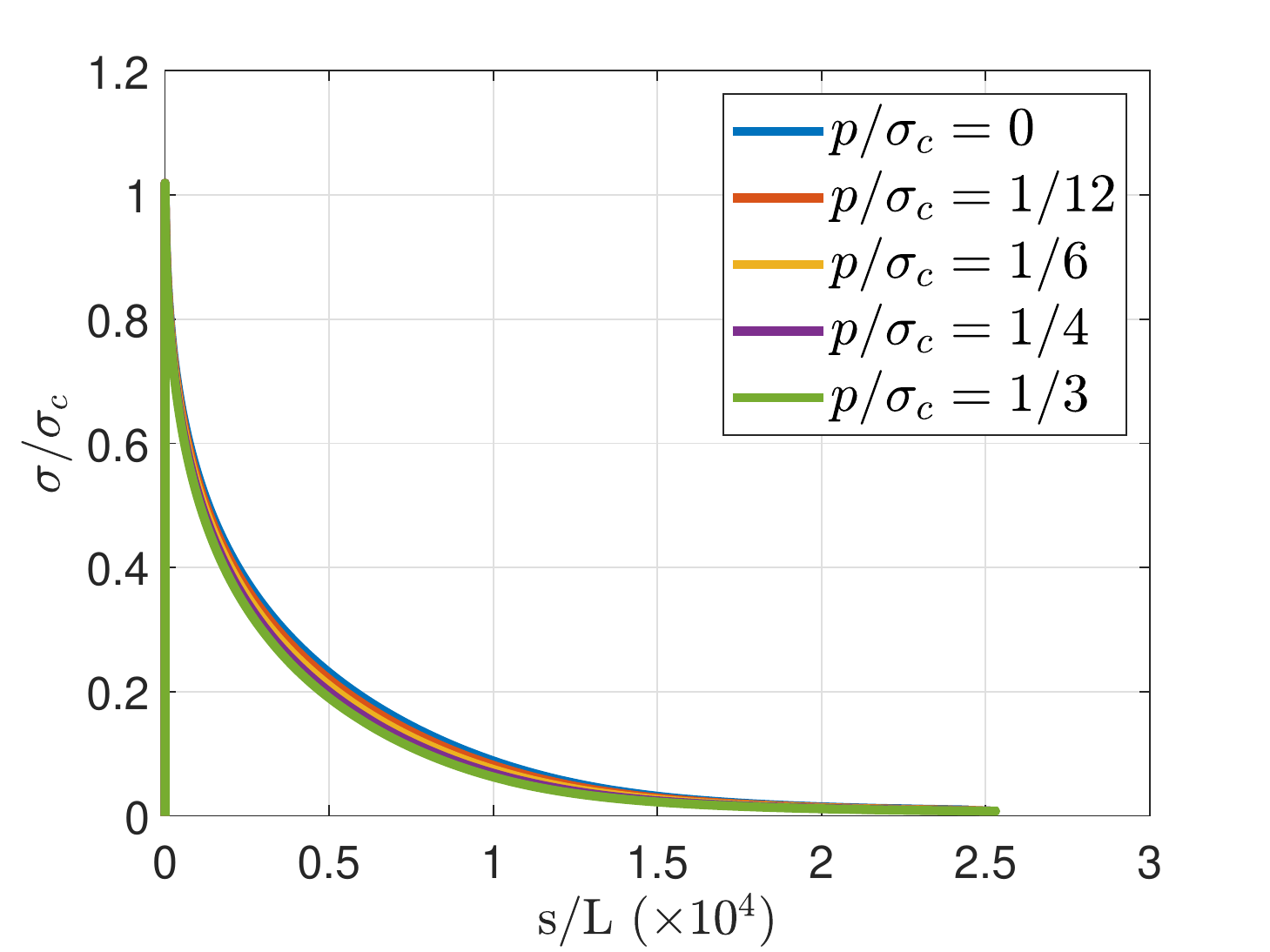}
  \caption{}
  \label{fig:traction_separation_bourdin_2d}
\end{subfigure}
\begin{subfigure}{.45\textwidth}
  \centering
  \includegraphics[width=\linewidth]{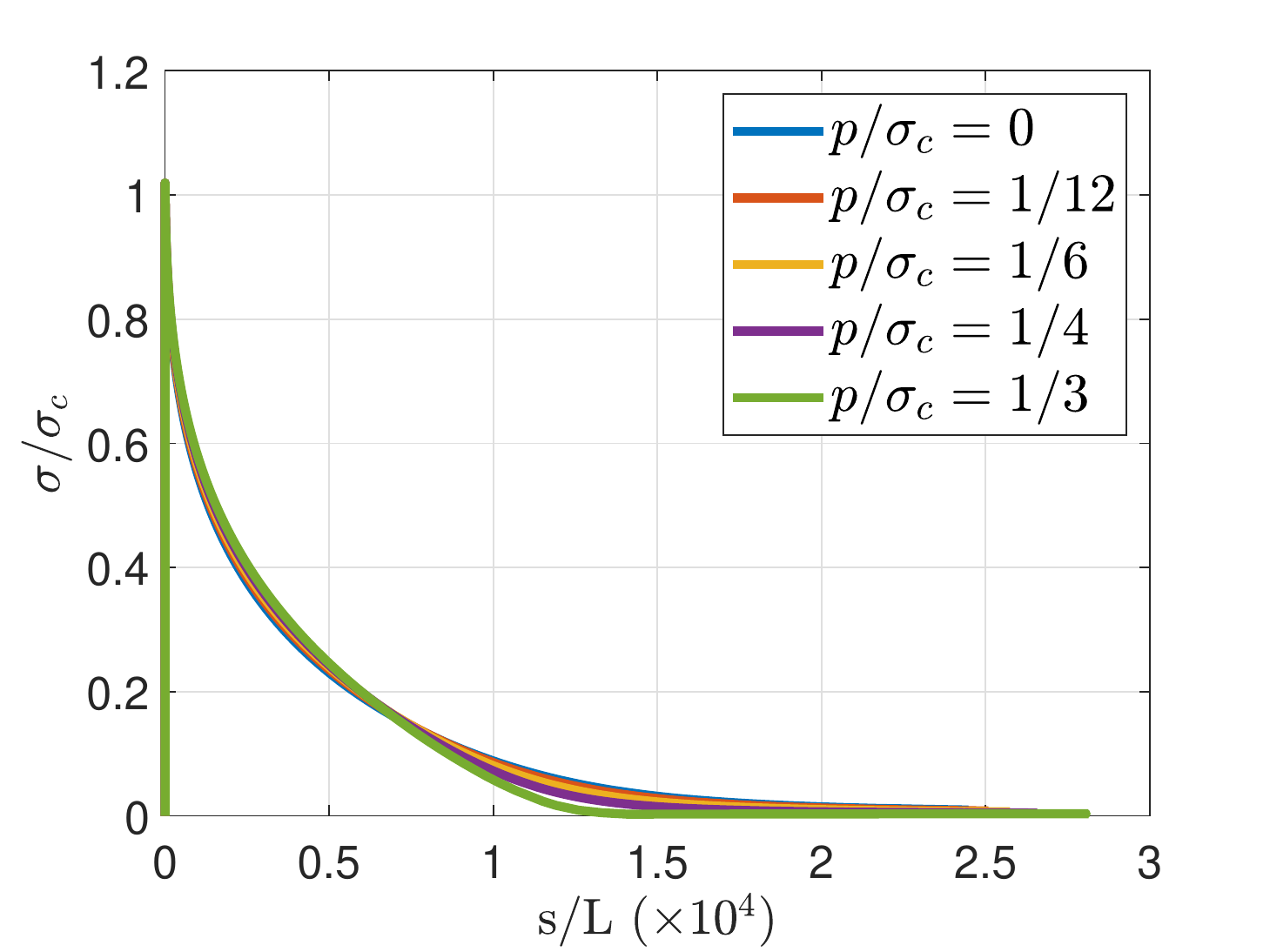}
  \caption{}
  \label{fig:traction_separation_gary}
\end{subfigure}
  \caption{Traction-separation curves for pressurized uniaxial cohesive bar problem, obtained with various phase-field models: (a) \eqref{lvc} with linear indicator function; (b) \eqref{lvc} with quadratic indicator function; (c) \eqref{lvc} with $2d-d^2$ indicator function; and (d) Proposed approach \eqref{uvc} with linear indicator function. } 
  \label{fig:traction_separation_results}
\end{figure*}

For the indicator function $I(d)$, results are reported for: (1) $I(d) = d$, used for example in \cite{bourdin2012variational}; (2) $I(d) = d^2$, used in \cite{jiang2022phase} and (3) $I(d) = 2d-d^2$, used in \cite{wheeler2014augmented}.
%post-processing details
The effective traction-separation laws extracted from the set of simulations are shown in Figure \ref{fig:traction_separation_results}. To generate these curves, the traction is computed as the internal force measured in the center of the bar. The separation $s$ is the opening of the crack, calculated as $s = -\int\limits_{-\infty}^{\infty}\textbf{u}\cdot\nabla I(d) \text{dx}$ \cite{bourdin2012variational}. 

%discussion of results
The results for the various models are shown in Figure \ref{fig:traction_separation_results}, with tractions and pressures normalized by the critical stress $\sigma_c$ from \eqref{eq:sigmacrit-from-psicrit}.
As shown in Figure \ref{fig:traction_separation_results}, the proposed model \eqref{uvc} exhibits minimal sensitivity to the pressure magnitude in the traction-separation behavior.  
By contrast, with the \eqref{lvc} formulation, only the case with $I(d) = 2d - d^2$ exhibits comparable results. In the other two cases (Figures \ref{fig:traction_separation_bourdin_d} and \ref{fig:traction_separation_bourdin_d2}), the apparent traction-separation law shows a spurious dependence to the applied pressure. This is evident in the variations in the results as well as the presence of jumps in the aperture at sufficiently high pressures. The latter occur due to an instability of the partially damaged solutions as $d$ approaches 1. More precisely, shortly after the damage at the center of the bar reaches $d\approx 0.8$, it jumps to $d= 1$, which in turns lead to a jump in the aperture.  
 This jump is indicated via the squares that appear on selected curves in Figures \ref{fig:traction_separation_bourdin_d} and \ref{fig:traction_separation_bourdin_d2}. The use of smaller displacement increments was not observed to significantly impact these results.   By contrast, such instabilities were not observed for the simulations reported in  Figures \ref{fig:traction_separation_bourdin_2d} and \ref{fig:traction_separation_gary}.

\FloatBarrier
\subsection{Crack nucleation from a pressurized hole}

\begin{figure*}[h]
% \centering
\begin{subfigure}{.49\textwidth}
  \centering
  \includegraphics[width=0.8\linewidth]{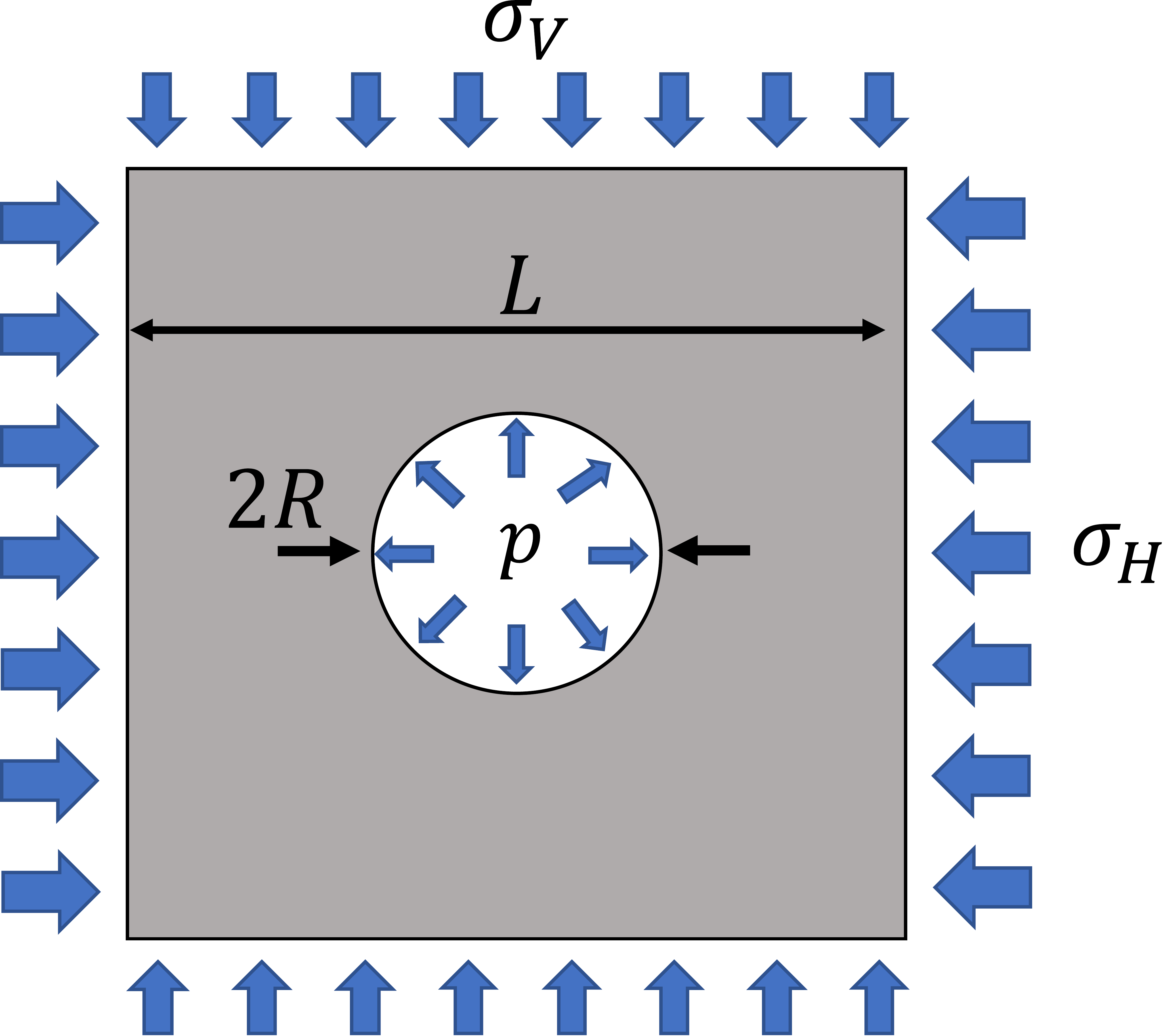}
  \caption{}
  \label{fig:cavity_schematic}
\end{subfigure}%
\begin{subfigure}{.49\textwidth}
  \centering
  \includegraphics[width=0.71\linewidth]{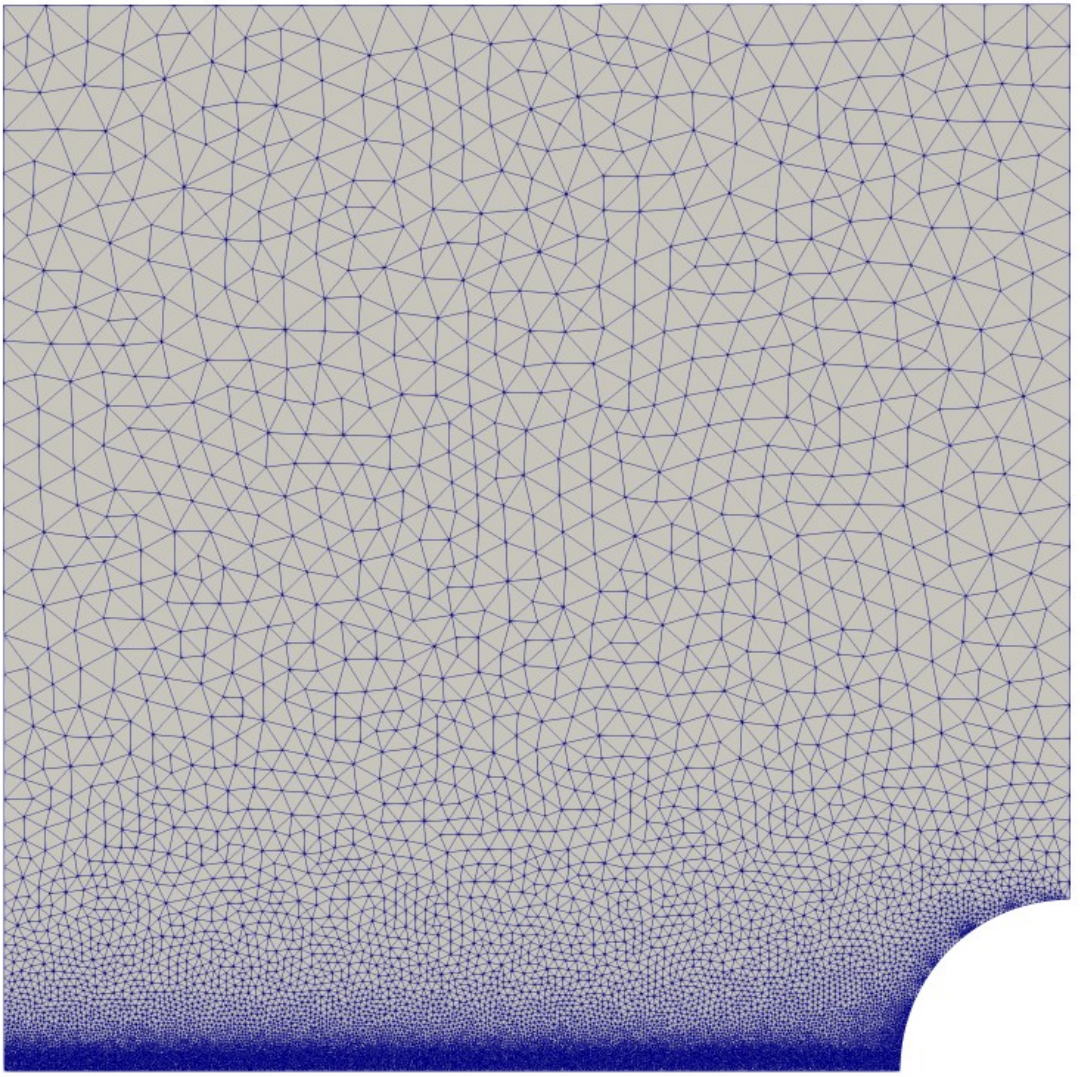}
  \caption{}
  \label{fig:quarter_mesh}
\end{subfigure}%
  \caption{(a) Problem schematic; (b) Mesh used in the computations, exploiting symmetry. } 
  \label{fig:initiation_problem_setup}
\end{figure*}

Consider a square plate of dimensions $L \times L$, with a circular hole in the center subjected to an internal pressure $p$, as shown in Figure \ref{fig:cavity_schematic}. This problem is motivated by oil and gas wellbore systems.   Far field stresses $\sigma_V$ and $\sigma_H$ are applied as tractions on the boundaries as shown. 
%This setup can be viewed, for example, as the cross-section of an oil and gas wellbore. 
The pressure is increased until it reaches a ``breakdown pressure" $p_b$. When that happens, cracks initiate in the direction parallel to the maximum \textit{in-situ} stress. Assuming $\sigma_H > \sigma_V$, this is expected to occur along a horizontal axis passing through the center of the hole.  In this work, the pressure in the hole is assumed to follow the crack faces as the fracture grows into the interior of the domain.  

\begin{table}[h]
\centering
\caption{Material properties, geometric parameters and applied loads for crack initiation problem}
\begin{tabular}[t]{lcc}
\hline
&Value &Unit \\
\hline
Young's modulus ($\text{E}$)&19.0$\times10^3$&MPa\\
Poisson's ratio ($\nu$)&0.2&--\\
Nucleation energy ($\psi_c$)&7.96$\times10^{-4}$&$\text{mJ mm}^{-3}$\\
Critical fracture energy  ($G_c$)&7.70$\times10^{-2}$&$\text{mJ mm}^{-2}$\\
Cavity radius ($R$)&400&mm\\
Specimen length ($L$)&5.0$\times10^{3}$&mm\\
Horizontal stress ($\sigma_H$)&5.0&MPa\\
Vertical stress ($\sigma_V$)&2.5&MPa\\
\hline
\end{tabular}
\label{material_properties_initiation}
\end{table}

The material properties selected for this problem, along with the dimensions and loading parameters are listed in Table \ref{material_properties_initiation}.  The material properties are taken to be representative of a Bebertal sandstone, as inspired by the experiments of \cite{stoeckhert2015fracture}.

The symmetry of the problem is exploited to reduce the computational domain to the top-left quarter.  An unstructured triangular mesh is used, with local refinement along the $x$-axis, as shown in Figure \ref{fig:quarter_mesh}. The element size in the refined area is $10$mm, whereas the phase-field regularization length is $\ell = 40$mm. For the results reported in this section, the phase-field model employs the cohesive formulation\cite{lorentz2011convergence, geelen2019phase} using the degradation function \eqref{cohesive_degradation} and the spectral split of  \cite{miehe2010phase}.  

Intuitively, the magnitude of the pressure load required to initiate fracture in this problem is expected to be independent of whether or not the pressure follows the crack evolution. After initiation, the pressure effects become important and the fracture propagates unstably. Due to this unstable behavior, it is very difficult to numerically capture the crack path after the pressure $p_b$ is reached. In order to have a glimpse into what this path looks like, a viscous term $\eta \dot d$ is added to the phase-field equation, as in \cite{miehe2010phase}, with $\eta = 10^{-3}\  \text{mJ}\cdot\text{mm}^{-3}\cdot$s. 

It bears emphasis that the equations \eqref{disp equation} and \eqref{damage equation} indicate that, in the absence of any damage, the proposed model for pressurized cracks reduces to the standard phase-field fracture model for traction-free cracks. Therefore, one should expect the proposed model to capture fracture initiation properly in this scenario. On the other hand, for the \eqref{lvc} formulation, this only occurs if the indicator function satisfies $I'(0) = 0$. Among the many works which use the \eqref{lvc} formulation, only a few such as \cite{jiang2022phase, peco2017influence} used an indicator function satisfying this condition. In \cite{jiang2022phase}, the authors were indeed able to predict fracture initiation from pressurized holes. To highlight the implications of having $I'(0) \neq 0$ in the model \eqref{lvc}, the results for this problem will also be presented using the  \eqref{lvc} formulation with the indicator function $I(d) = d$.

\begin{figure*}[h]
% \centering
\begin{subfigure}{.49\textwidth}
  \centering
  \includegraphics[width=0.7\linewidth]{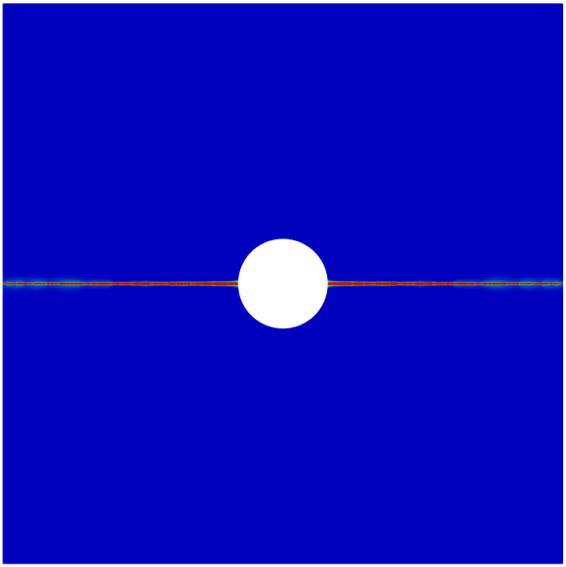}
  \caption{}
  \label{fig:damage_profile_gary}
\end{subfigure}%
\begin{subfigure}{.49\textwidth}
  \centering
  \includegraphics[width=0.86\linewidth]{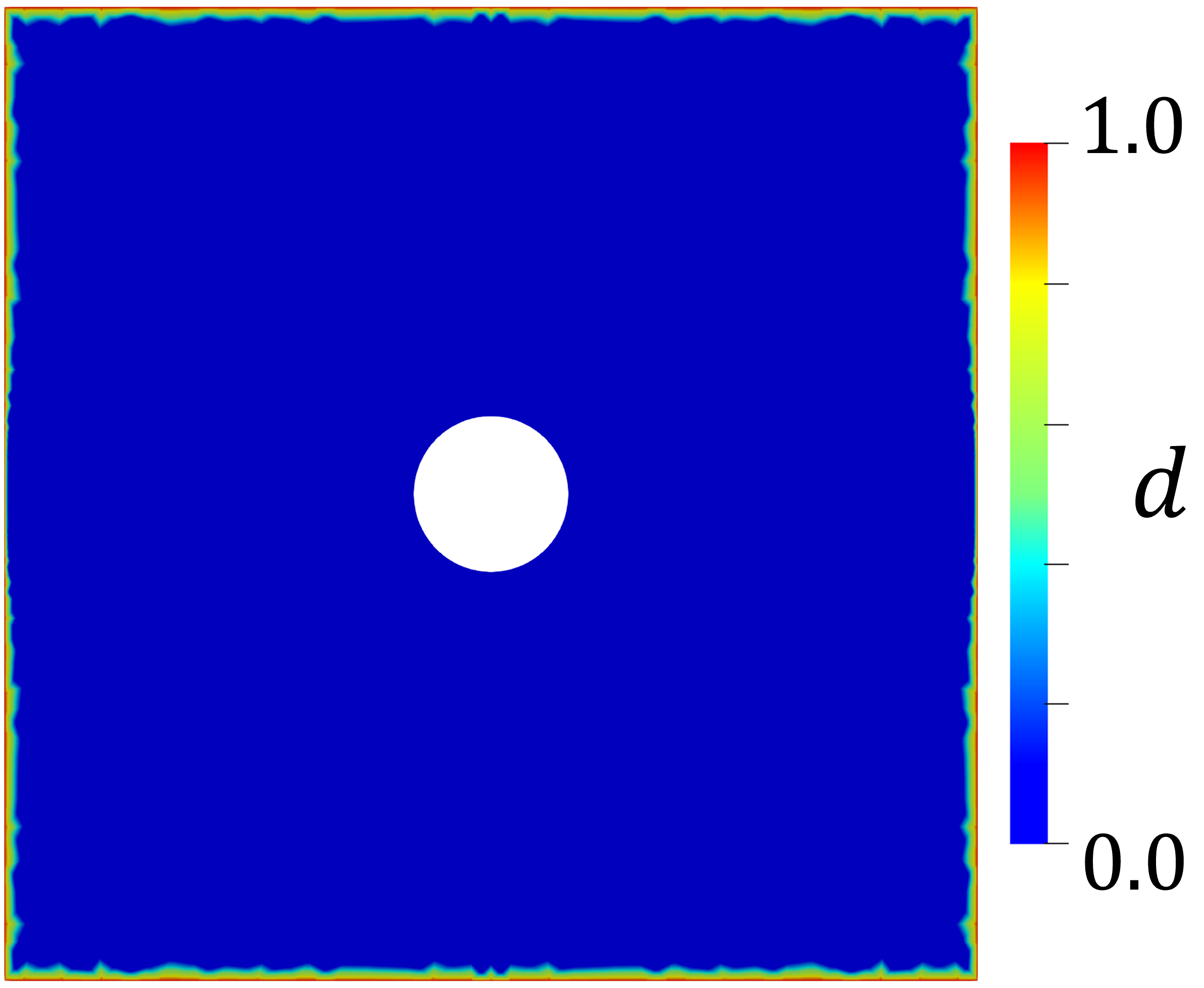}
  \caption{}
  \label{fig:damage_profile_bourdin}
\end{subfigure}%
  \caption{(a) Final crack pattern using proposed model; (b) Damage field using the model from \cite{bourdin2012variational}. } 
  \label{fig:damage_profiles}
\end{figure*}

The final damage patterns obtained using the \eqref{uvc} formulation and the \eqref{lvc} formulation  are shown in Figure \ref{fig:damage_profiles}. With the \eqref{uvc} formulation, damage localizes along the midplane when the hoop stress is approximately 85\% of $\sigma_c$.  This is not unexpected, as the expression \eqref{eq:sigmacrit-from-psicrit} is based on a one-dimensional state of stress and strain which differs significantly from the state near the corner of the hole.    The same comparison is not performed for the simulation using the model \eqref{lvc}, since damage forms only on the boundary in the first steps leading to spurious rigid body motion. 

The main takeaway is that the proposed model \eqref{uvc} allows one to study crack nucleation and subsequent propagation under a pressure load, whereas formulation \eqref{lvc} leads to spurious damage formation if $I'(0) \neq 0$. The presence of the term $p\nabla \cdot \textbf{u}I'(d)$ in the damage equation \eqref{wet damage equation box2} drives crack formation in areas which are not stressed. For this specific problem, this issue can be circumvented using for example $I(d) = d^2$, as shown in \cite{jiang2022phase}, but this option introduces a spurious dependence of the cohesive response of the material on the applied pressure, as indicated in the last section (Figure \ref{fig:traction_separation_bourdin_d2}).

\subsection{Stable propagation of a pre-existing crack}

\begin{figure*}[h]
% \centering
\begin{subfigure}{.49\textwidth}
  \centering
  \includegraphics[width=0.8\linewidth]{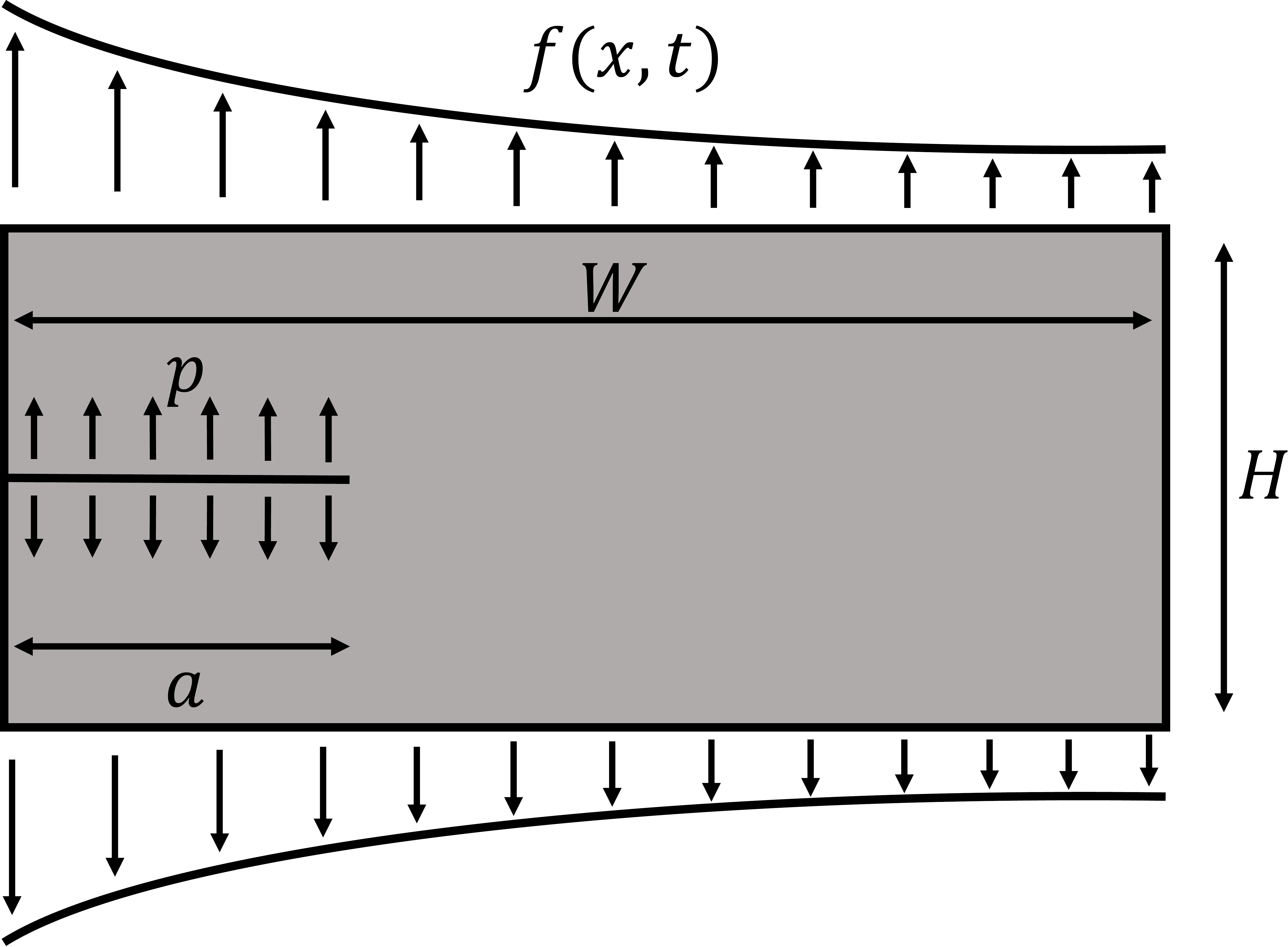}
  \caption{}
  \label{fig:surfing_schematic}
\end{subfigure}%
\begin{subfigure}{.49\textwidth}
  \centering
  \vspace{1.06cm}
  \includegraphics[width=0.8\linewidth]{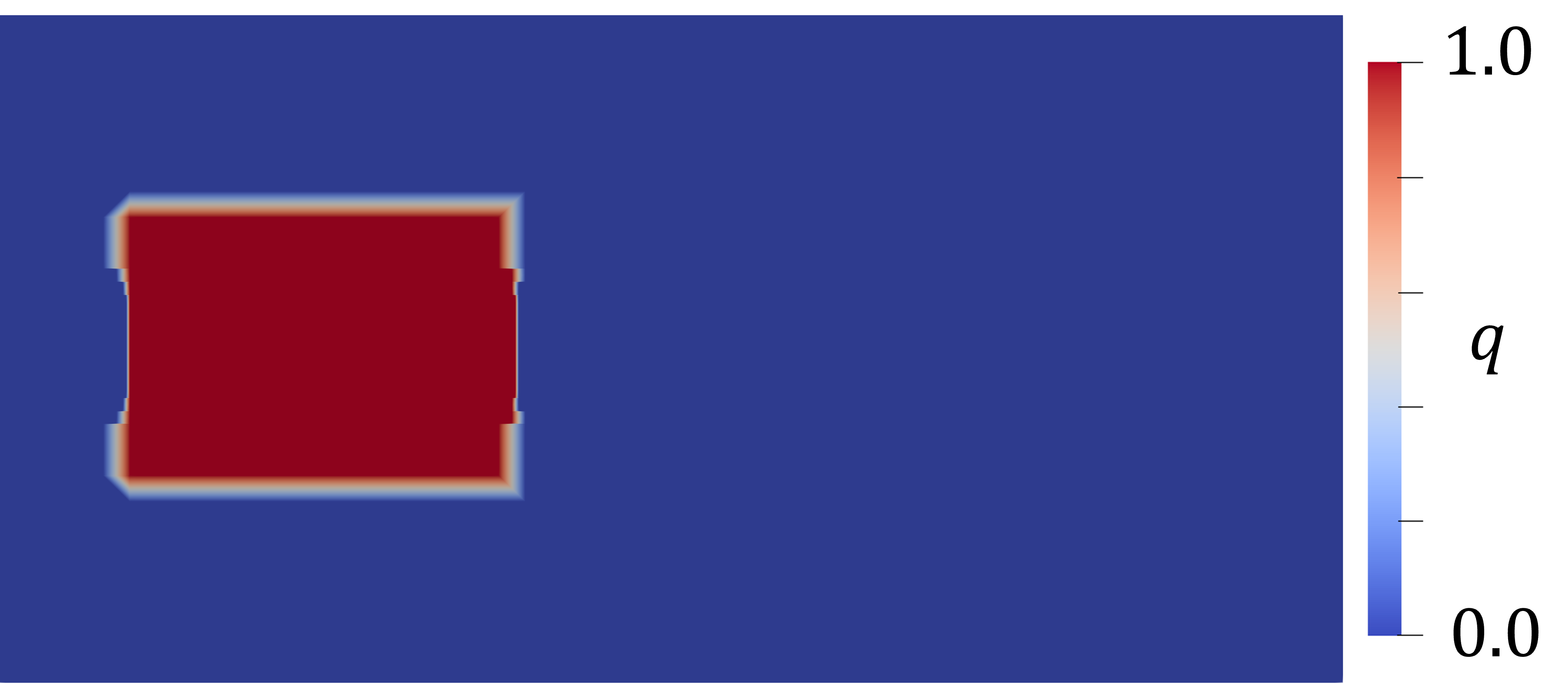}
  \vspace{1.06cm}
  \caption{}
  \label{fig:integration_domain}
\end{subfigure}%
  \caption{(a) Geometry and boundary conditions for pressurized crack propagation problem; (b) J-Integral domain function $q$. } 
  \label{fig:surfing_problem_setup}
\end{figure*}

Consider a strip of material with a pressurized crack, as shown in  Figure \ref{fig:surfing_schematic}.
The rectangular strip has a width $W$, height $H$ and a crack with initial size $a$ (values provided in Table \ref{material_properties_propagation}), and is loaded by the ``surfing" boundary condition $\widetilde{U_y}(x,y,t)$ on its top and bottom surfaces \cite{hossain2014effective}. The boundary condition is given by
\begin{equation}\label{surfing_bc_1}
    \widetilde{U_y}(x,y,t) = U_y(x-Vt,y),  
\end{equation}
where
\begin{equation}\label{surfing_bc_2}
    U_y(x,y) = \hat{U}_y(r,\theta) = \dfrac{\sqrt{G_cE'}}{2\mu}\sqrt{\dfrac{r}{2\pi}}(\kappa - \cos\theta)\sin\dfrac{\theta}{2},
\end{equation}
and where $r$ and $\theta$ are polar coordinates with respect to the origin, taken to be the midpoint of the left edge of the domain. The constant $V>0$ is the target crack speed, prescribed by moving the boundary condition following \eqref{surfing_bc_1}. The Kolosov constant is defined as $\kappa = 3-4\nu$ in plane strain and the shear modulus $\mu = E/(2+2\nu)$. The pressure $p$ applied to the crack faces as the crack evolves is given by
\begin{equation}
    p = \dfrac{1}{2}\sqrt{\dfrac{G_cE'}{\pi a}},
\end{equation}
 in which $a$ denotes the initial crack length. This value corresponds to half the critical pressure for an infinite plate with a pressurized crack of size $a$. This magnitude ensures that the applied pressure is considerably large, but not so large as to drive the problem beyond the stable propagation regime. 
 
 To calculate the energy release rate, the domain form of the J-integral \eqref{j_integral_theorem} developed in Section \ref{sec:j_integral} is used. The function $q$ is constructed by taking advantage of the finite element interpolation. In essence, the domain for the J-integral is taken to be a single rectangular region of dimensions $a \times H/2$, centered on the initial crack tip. The value of $q$ for all nodes outside of this region is set to $0$, while $q = 1$ for all nodes inside. Using the finite element interpolation, this gives rise to a $q$ function whose value changes continuously from $0$ to $1$ on the elements cut by the rectangular path. This function is illustrated in Figure \ref{fig:integration_domain}\footnote{Due to mesh refinement near the crack surface, the width of the band where $0 < q < 1$ diminishes near the horizontal centerline of the domain.}.

\begin{table}[h]
\centering
\caption{Parameters used for pressurized crack propagation problem}
\begin{tabular}[t]{lcc}
\hline
&Value &Unit \\
\hline
Young's modulus ($\text{E}$)&3.0$\times10^4$&MPa\\
Poisson's ratio ($\nu$)&0.2&--\\
Critical fracture energy  ($G_c$)&0.12&$\text{mJ mm}^{-2}$\\
Initial crack length ($a$)&1.6&m\\
Specimen width ($W$)&8.0&m\\
Specimen height ($H$)&4.0&m\\
Target crack speed ($V$)&0.4&m/s\\
\hline
\end{tabular}\label{material_properties_propagation}
\end{table}

In order to verify that Griffith's law is approached as $\ell \rightarrow 0$, simulations are performed for this problem using a sequence of decreasing regularization lengths, ranging from $\ell = a/20$ to $\ell = a/160$. The mesh is locally refined along the $x$-axis, where the element size is set to $h = \ell/4$. The symmetry of the problem is exploited and only the response in the top half of the domain is simulated. 

In terms of constitutive choices of the phase-field model, the AT-1 formulation is employed without any decomposition of the strain. 
 
As in the previous examples, this problem is analyzed using the formulations \eqref{lvc} and \eqref{uvc}, and the following choices of indicator function $I(d)$:
\begin{itemize}
    \item $I(d) = d$
    \item $I(d) = d^2$
    \item $I(d) = 2d-d^2$ 
\end{itemize}

\begin{figure*}[h]
% \centering
\begin{subfigure}{.33\textwidth}
  \centering
  \includegraphics[width=\linewidth]{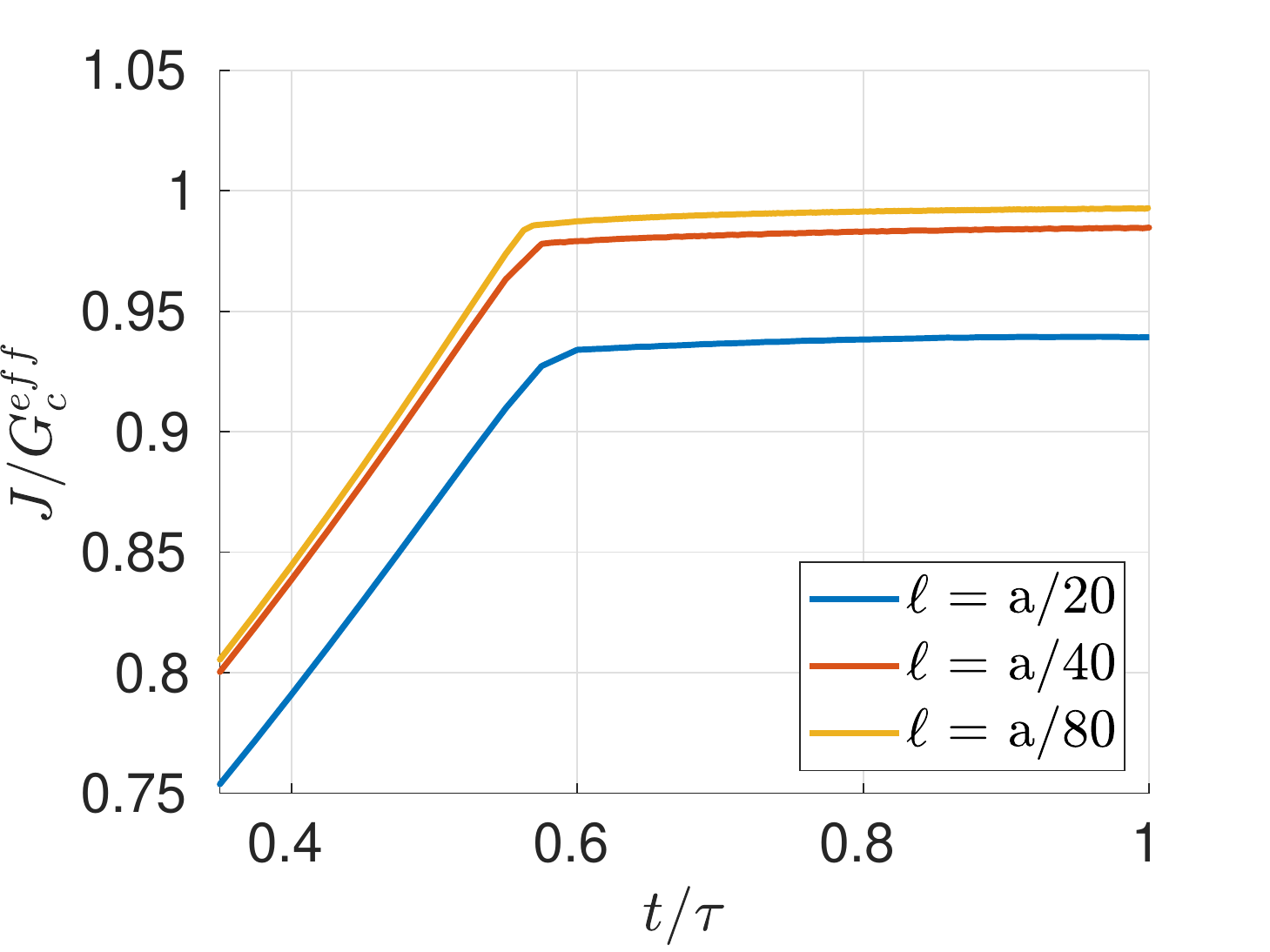}
  \caption{}
  \label{fig:prop_bourdin_d}
\end{subfigure}%
\begin{subfigure}{.33\textwidth}
  \centering
  \includegraphics[width=\linewidth]{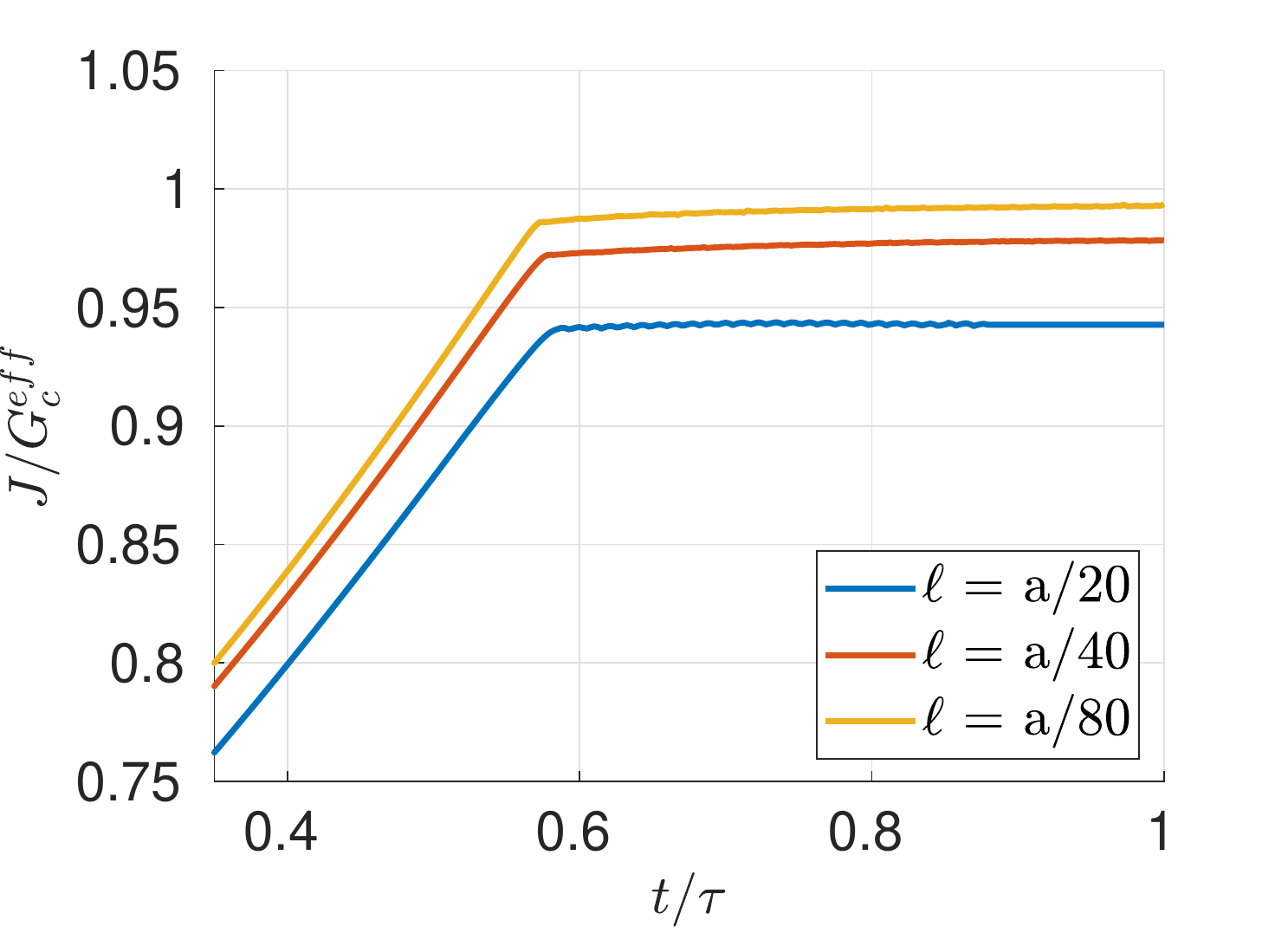}
  \caption{}
  \label{fig:fig:prop_bourdin_d2}
\end{subfigure}%
\begin{subfigure}{.33\textwidth}
  \centering
  \includegraphics[width=\linewidth]{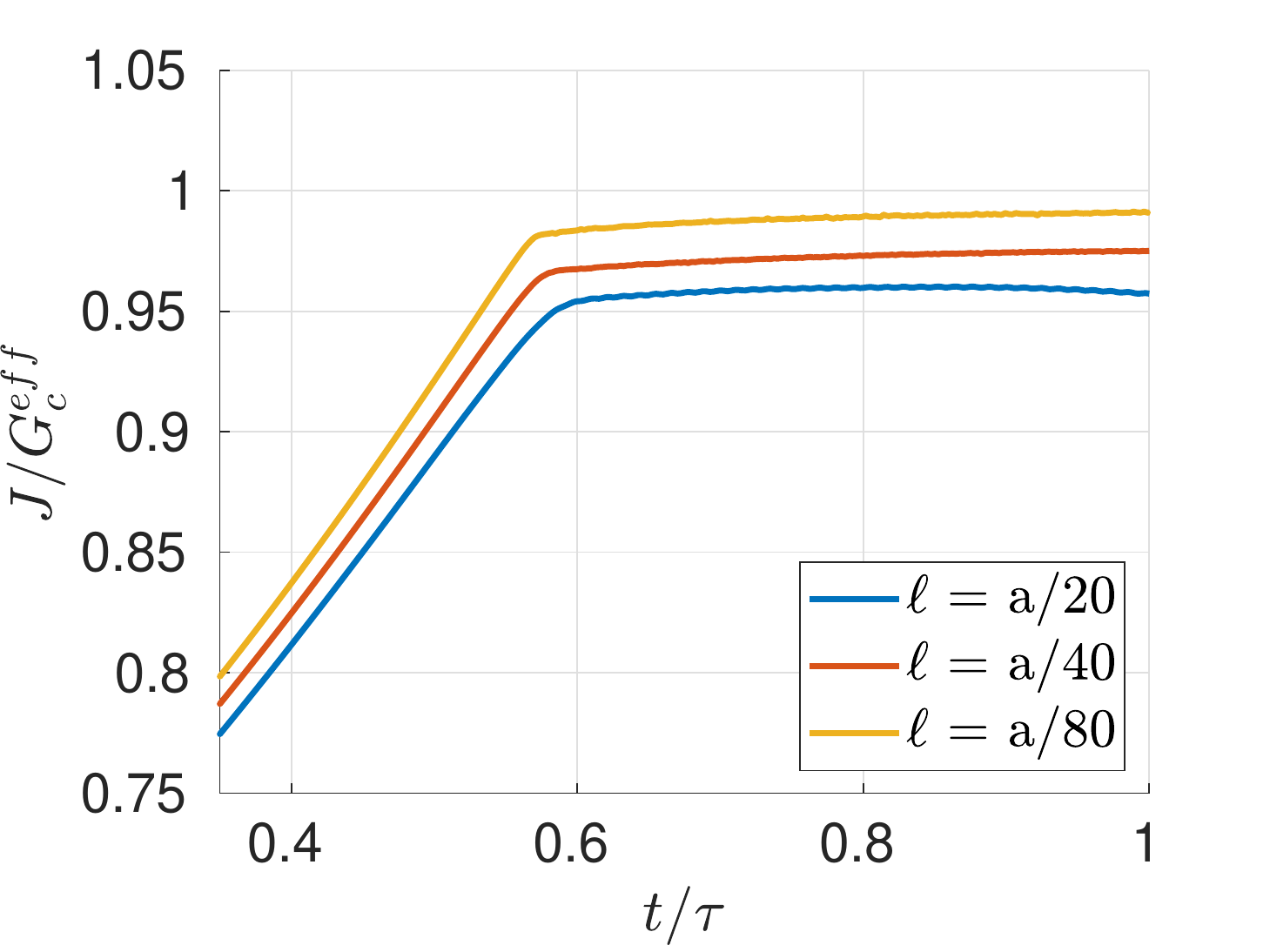}
  \caption{}
  \label{fig:prop_bourdin_2d}
\end{subfigure}
  \caption{Reference results with the \ref{lvc} formulation. Curves with $\ell = a/160$ are not shown, as they are almost identical to the ones with $\ell = a/80$. (a) $I(d) = d$; (b) $I(d) = d^2$; (c) $I(d) = 2d-d^2$  } 
  \label{fig:prop_bourdin}
\end{figure*}

\begin{figure*}[h]
% \centering
\begin{subfigure}{.33\textwidth}
  \centering
  \includegraphics[width=\linewidth]{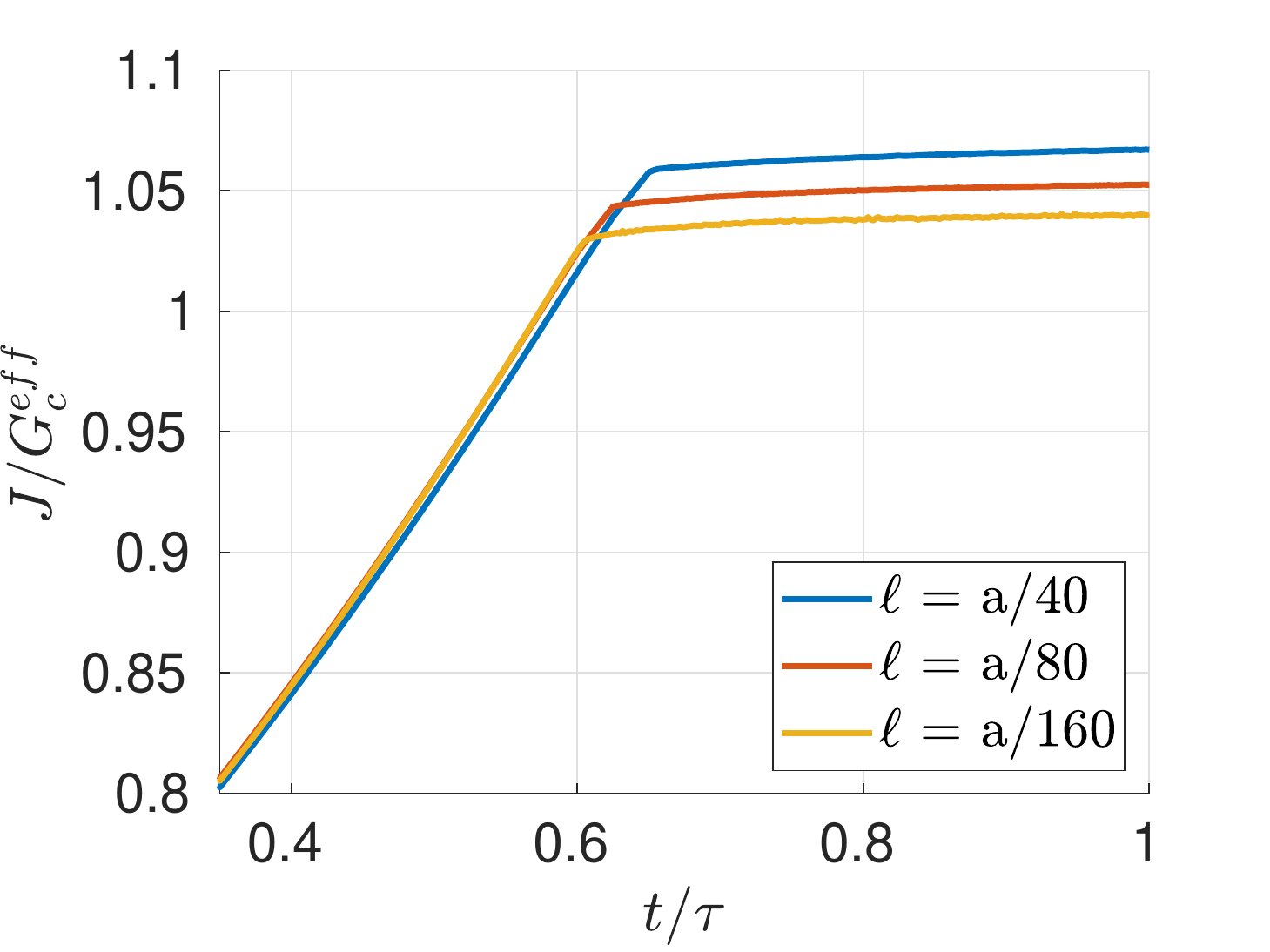}
  \caption{}
  \label{fig:prop_gary_d}
\end{subfigure}%
\begin{subfigure}{.33\textwidth}
  \centering
  \includegraphics[width=\linewidth]{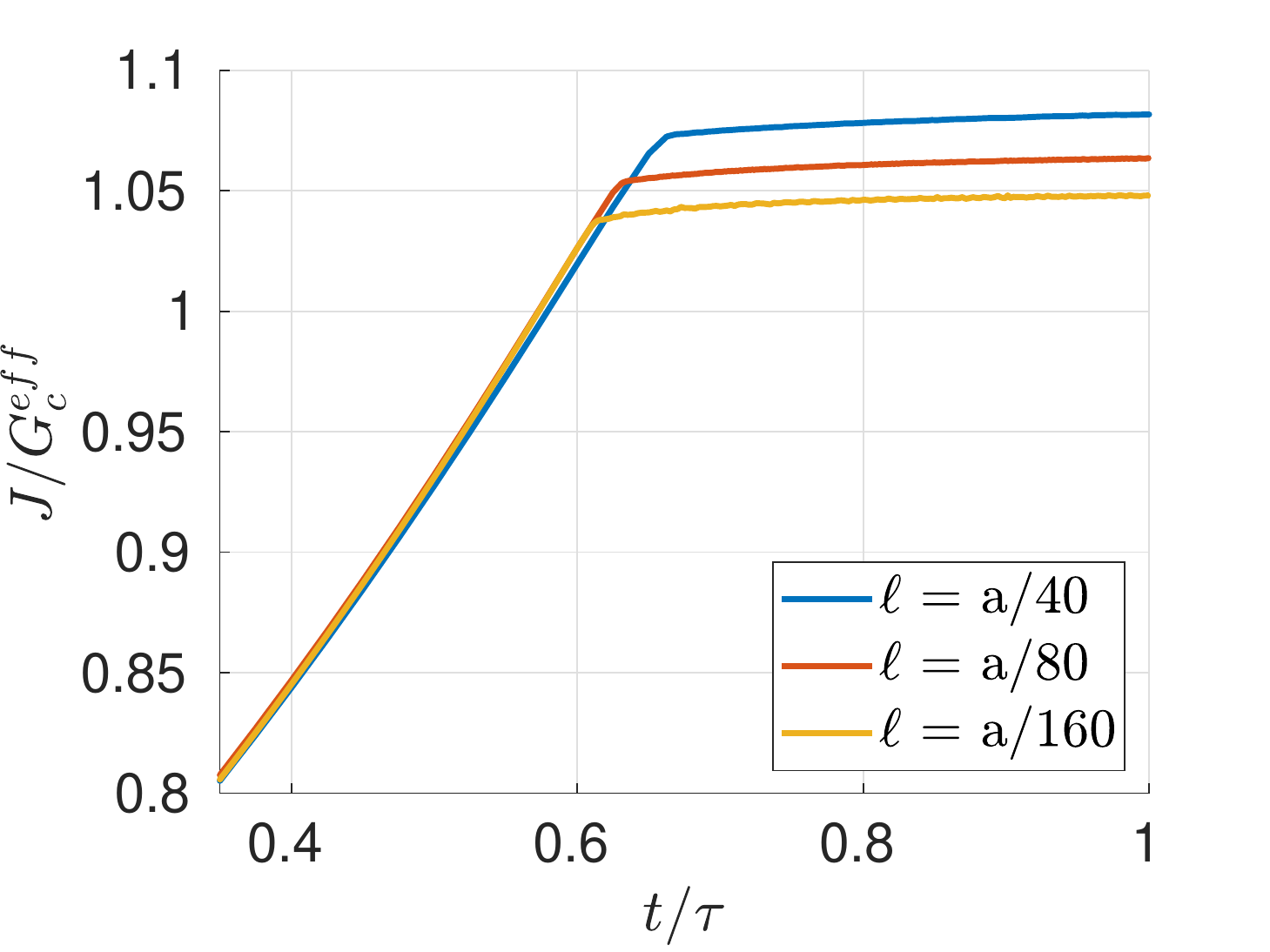}
  \caption{}
  \label{fig:prop_gary_d2}
\end{subfigure}%
\begin{subfigure}{.33\textwidth}
  \centering
  \includegraphics[width=\linewidth]{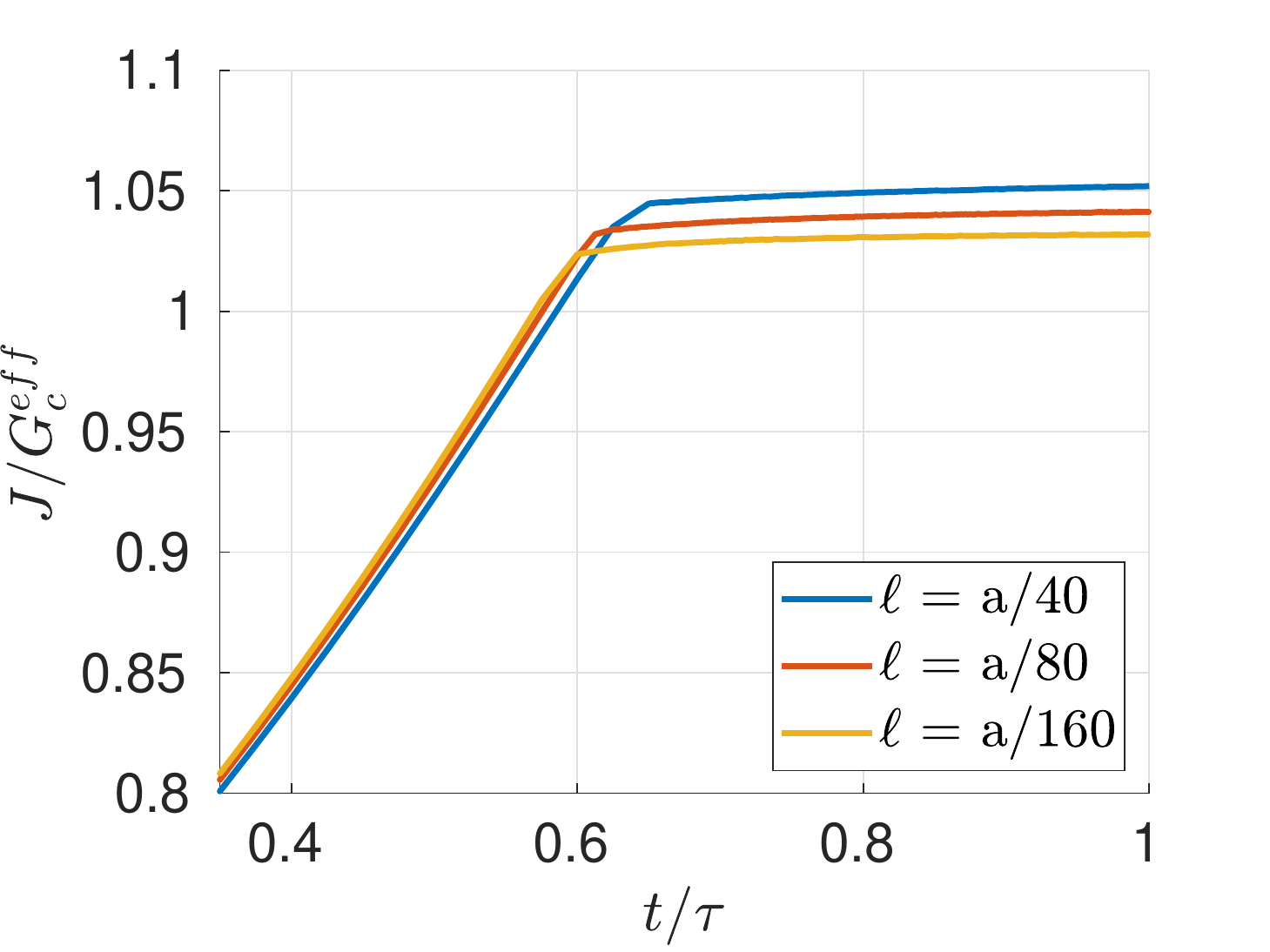}
  \caption{}
  \label{fig:prop_gary_2d}
\end{subfigure}
  \caption{Results with the proposed formulation \eqref{uvc} (a) $I(d) = d$; (b) $I(d) = d^2$; (c) $I(d) = 2d-d^2$  } 
  \label{fig:prop_gary}
\end{figure*}

To evaluate how well the models approach Griffith's law, the ratio between the energy release rate measured by the J-Integral and the effective critical fracture energy $G^{eff}_c = (1+2h/c_0\ell) G_c$ \footnote{in fact, phase-field cracks actually dissipated a slightly larger energy per unit length in numerical models. A correction factor of $\left(1+\dfrac{2h}{c_0\ell}\right)$ is then applied to $G_c$, following \cite{yoshioka2020crack}. The factor of 2 here comes from the symmetry boundary condition employed.} is plotted in Figures \ref{fig:prop_bourdin} and \ref{fig:prop_gary}.  In all figures, the time is scaled by a characteristic time $\tau$, defined as $\tau = a/V$. %Since the loading stage before the onset of propagation is not particularly important in this example, most of the that part ($t < 0.3\tau$) is omitted for better visualization of the results during the propagation phase.
The results using the traditional \eqref{lvc} formulation are presented in Figure \ref{fig:prop_bourdin}. They indicate convergence towards $J/G^{eff}_c$ = 1 as the regularization length is reduced, especially when the indicator function $I(d)=d$ is used. This is expected given the results obtained in \cite{bourdin2012variational}. Nevertheless, these results serve to verify the implementation of the J-Integral presented in Section \ref{sec:j_integral}. They also provide an estimate for how small the regularization length has to be in order to achieve a certain level of accuracy with these types of phase-field models.

\begin{figure}
      \centering
      \includegraphics[width=\linewidth]{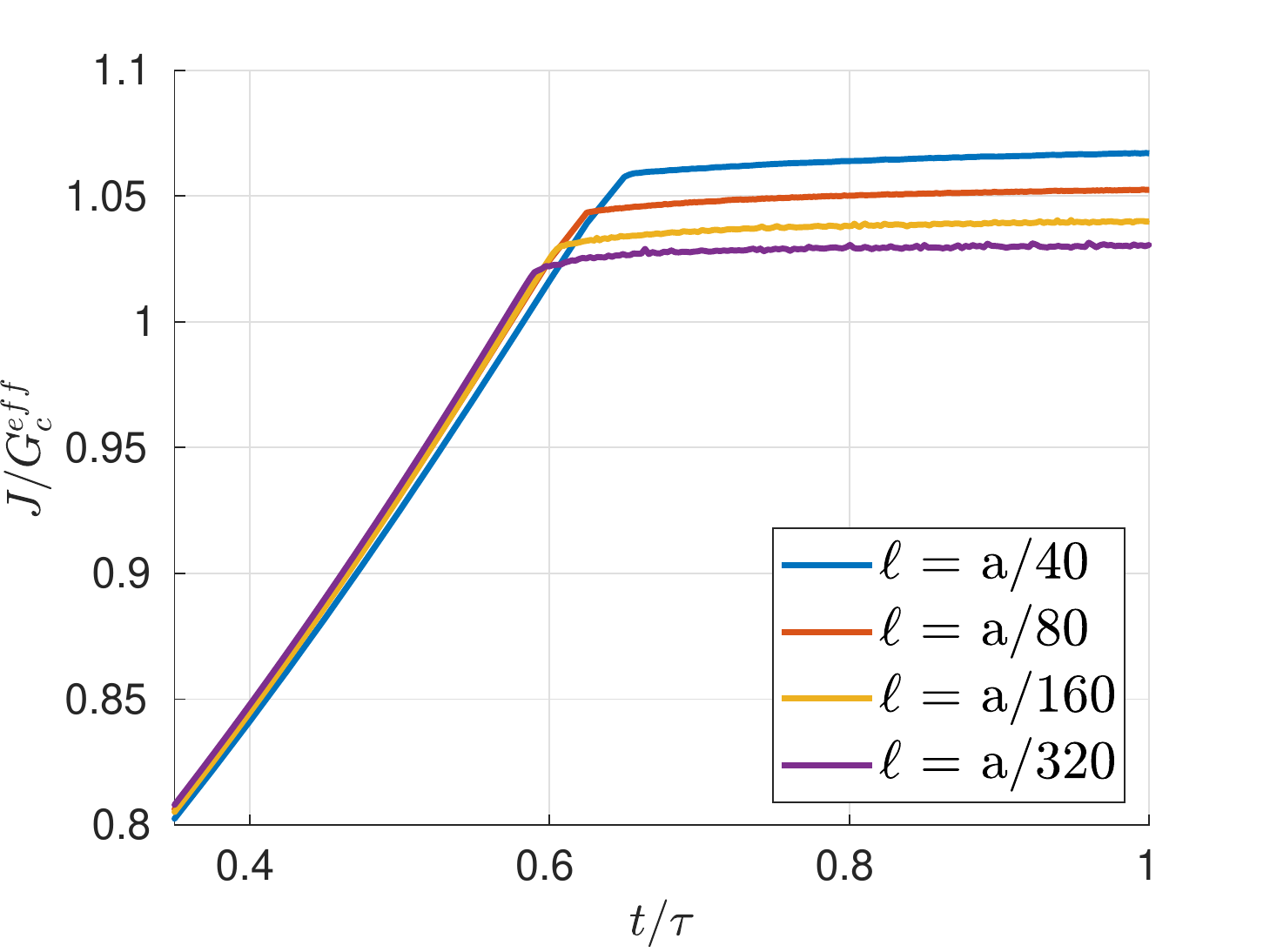}
      \captionof{figure}{Convergence of the proposed formulation with $I(d) = d$.}
      \label{fig:refined_gary_I_d}
\end{figure}

\begin{table}[h]
    \centering
    \begin{tabular}{ccc}
    \hline
    $\ell/a$ &Error &$\text{Error}_{k+1}/\text{Error}_{k}$ \\
    \hline
    1/40  & 0.067 & --\\
    1/80  & 0.052 & 0.78\\
    1/160 & 0.040 & 0.77\\
    1/320 & 0.031 & 0.77\\
    \hline
    \end{tabular}
    \caption{Absolute error in $J$ vs.\ $G_c^{eff}$ for the pressurized crack propagation problem, as a function of regularization length.}
    \label{table:convergence_check}
\end{table}

For the case of proposed formulation \eqref{uvc}, the results shown in Figure \ref{fig:prop_gary} indicate a slower convergence towards a 
$J/G^{eff}_c = 1$ response. In contrast with the \eqref{lvc} formulation, the curves converge from above, and therefore, the fracture toughness is slightly overestimated when larger regularization lengths are used. Nevertheless, they all seem to approach a Griffith-like response in the limit $\ell \rightarrow 0$. In Figure \ref{fig:refined_gary_I_d}, an even finer result, using \eqref{uvc} with $I(d)=d$ and $\ell = a/320$ is added, to ensure that the convergence rates indicated in Figure \ref{fig:prop_gary} persist. In Table \ref{table:convergence_check}, the relative errors are provided, indicating a convergence rate of approximately $0.4$ with respect to $\ell$.

One potential explanation for the slower convergence rate is related to the different assumptions regarding the trial cracks, as discussed in Section \ref{sec:model}. Although the different assumptions converge to the same propagation rule in the limit of an infinitesimal crack increment, in the discretized case, the minimal crack increment is finite and related to the mesh spacing $h$ and regularization length $\ell$. In this case, a slightly different propagation behavior, resulting in slower convergence rates towards $J/G^{eff}_c = 1$ is not surprising.

\section{Concluding Remarks}

This manuscript examines various models for phase-field fracture incorporating pressure loads on diffuse crack faces.  This includes the analysis of a new formulation that can be obtained by considering the presence of the pressure load in the virtual extension of a crack, or alternatively through a careful accounting in the minimization procedure. The new formulation is referred to as the 
 ``unloaded virtual crack formulation"\eqref{uvc}. In order to verify the accuracy of the various models for propagating cracks, a new form of the J-Integral for pressurized cracks in the phase-field context is derived.

The \eqref{uvc} formulation proposed herein allows for a unified treatment of crack nucleation and propagation in scenarios involving either brittle or cohesive fracture, and provides for better accuracy in some problems compared to existing formulations of the \eqref{lvc} type. As it allows for the use of the same governing equation for the damage parameter, its computational implementation within existing phase-field solvers is also simpler. In future work, its applicability to problems involving plastic deformation and strength-based fracture nucleation will be studied. In addition to that, modifications to accelerate the convergence of the model with respect to the phase-field parameter $\ell$ will also be considered.

\section{Acknowledgments}

%The authors would like to acknowledge the support and funding from Duke University and Argonne National Laboratory, which was essential to the development of this work. 
The partial support of A.\ Costa and  J.E.\ Dolbow by the National Science Foundation, through grant CMMI-1933367 to Duke University, is gratefully acknowledged.  
 T.\ Hu gratefully acknowledges the support of Argonne National Laboratory.  Argonne National Laboratory is managed and operated by UChicago Argonne LLC.\ for the U.S.\ Department of Energy under Contract No.\ DE-AC02-06CH11357.

\newpage

\appendix
\section{Equivalence to SIF condition}\label{SIF_equivalence}

In this appendix, the energy release rate for a single, straight crack under an arbitrary pressure load $p(x)$ is computed, assuming that infinitesimal crack increments are traction free, as shown in Figure \ref{fig:dry_crack}. Griffith's criteria states that propagation should happen whenever this energy release rate, which will be denoted by $G$, reaches $G_c$. This will give rise to a condition for propagation based on the pressure distribution $p(x)$, the crack size $a$ and Young's modulus $E'$\footnote{assuming plane strain, $E'=E/(1-\nu^2)$}. The purpose of the following derivation is to demonstrate that this condition is equivalent to the stress intensity factor criteria \cite{irwin1957analysis}.

Initially, consider the Sneddon-Lowengrub solution for the aperture of a pressure loaded crack in an infinite plate, under plane strain conditions,

\begin{equation}
    w(x) = \dfrac{4a}{\pi E'}\int^1_0p(sa)Z(x/a, s)\text{d}s
\end{equation}

\noindent where

\begin{equation}
    Z(r,s) = \log\left|\dfrac{\sqrt{1-r^2}+\sqrt{1-s^2}}{\sqrt{1-r^2}-\sqrt{1-s^2}}\right|
\end{equation}

\noindent is a convolution kernel. The work done by the pressure load is then,

\begin{equation}
    W_p = \int_{-a}^a p w \text{d}x = \int_{-a}^a p(y) \dfrac{4a}{\pi E'}\int^1_0p(sa)Z(x/a, s)\text{d}s\text{d}y.
\end{equation}

\noindent Clayperon's theorem \cite{fosdick2003} states that the potential energy is negative half of the work exerted in the boundary, which, in this case is only $W_p$. Hence

\begin{multline}
    U = -\dfrac{1}{2}W_p = -\dfrac{1}{2}\int_{-a}^a p w \text{d}x \\ = -\dfrac{4a^2}{\pi E'}\int_{0}^1 p(ar)\int^1_0p(sa)Z(r, s)\text{d}s\text{d}r.
\end{multline}

\noindent Let's write the energy release rate, assuming that the pressure field doesn't vary as the crack advances by a small amount $\text{d}a$. That is,

\begin{equation}
    p^{a+\text{d}a}(x) = 
    \begin{cases}
      p^{a}(x), & \text{if}\ x\le a \\
      0, & \text{if}\ a \le x \le a+\text{d}a
    \end{cases} 
\end{equation}

\begin{equation}
    dU = U(a+\text{d}a, p^{a+\text{d}a}) - U(a+\text{d}a, p^{a}), 
\end{equation}

\small{
\begin{multline}
       dU = \\  -\dfrac{4}{\pi E'}\int_{0}^{a+\text{d}a} p^{a+\text{d}a}(x)\int_0^{a+\text{d}a} p^{a+\text{d}a}(y)Z(\frac{x}{a+\text{d}a}, \frac{y}{a+\text{d}a}) \text{d}y\text{d}x \\ + \dfrac{4}{\pi E'}\int_{0}^a p^{a}(x)\int_0^a p^{a}(y)Z(x/a, y/a) \text{d}y\text{d}x.
\end{multline}
}
\normalsize

\noindent Using the definition of $p^{a+\text{d}a}$ given above,

\small{
\begin{multline*}
        dU = -\frac{4}{\pi E'} \times \\ \int_{0}^{a} p^{a}(x)\int_0^{a} p^{a}(y)\left(Z(\frac{x}{a+\text{d}a}, \frac{y}{a+\text{d}a}) - Z(x/a, y/a)\right)\text{d}y\text{d}x.
\end{multline*}
}
\normalsize

\noindent By symmetry, both tips of the crack propagate with the same energy release rate, so, one can write,

\small{
\begin{multline*}
    2G = -\dfrac{dU}{\text{d}a} = \dfrac{4}{\pi E'} \times \dfrac{1}{\text{d}a} \times \\ \int_{0}^{a} p^{a}(x)\int_0^{a} p^{a}(y)\left(Z(\frac{x}{a+\text{d}a}, \frac{y}{a+\text{d}a}) - Z(x/a, y/a)\right)\text{d}y\text{d}x.
\end{multline*}
}
\normalsize

\noindent The term between parenthesis can be re-written as,

\begin{multline}\label{Z_difference}
    Z(\frac{x}{a+\text{d}a}, \frac{y}{a+\text{d}a}) - Z(x/a, y/a) = \\ \log\left|\dfrac{\sqrt{(a+\text{d}a)^2-x^2}+\sqrt{(a+\text{d}a)^2-y^2}}{\sqrt{(a+\text{d}a)^2-x^2}-\sqrt{(a+\text{d}a)^2-y^2}}\right| \\ - 
    \log\left|\dfrac{\sqrt{a^2-x^2}+\sqrt{a^2-y^2}}{\sqrt{a^2-x^2}-\sqrt{a^2-y^2}}\right|\\
    = \log\left|\dfrac{\sqrt{(a+\text{d}a)^2-x^2}+\sqrt{(a+\text{d}a)^2-y^2}}{\sqrt{a^2-x^2}+\sqrt{a^2-y^2}}\right| \\ - 
    \log\left|\dfrac{\sqrt{(a+\text{d}a)^2-x^2}-\sqrt{(a+\text{d}a)^2-y^2}}{\sqrt{a^2-x^2}-\sqrt{a^2-y^2}}\right|.
\end{multline}

\noindent The second term contains a singularity, which can be removed if one re-writes it as,

\begin{multline}
    \log\left|\dfrac{\sqrt{(a+\text{d}a)^2-x^2}-\sqrt{(a+\text{d}a)^2-y^2}}{\sqrt{a^2-x^2}-\sqrt{a^2-y^2}}\right| = \\
    \log\biggl|\dfrac{\sqrt{(a+\text{d}a)^2-x^2}-\sqrt{(a+\text{d}a)^2-y^2}}{\sqrt{a^2-x^2}-\sqrt{a^2-y^2}} \\ \times \dfrac{\sqrt{(a+\text{d}a)^2-x^2}+\sqrt{(a+\text{d}a)^2-y^2}}{\sqrt{a^2-x^2}+\sqrt{a^2-y^2}} \\ \times \dfrac{\sqrt{a^2-x^2}+\sqrt{a^2-y^2}}{\sqrt{(a+\text{d}a)^2-x^2}+\sqrt{(a+\text{d}a)^2-y^2}} \biggr| \\
    = \log\biggl|\dfrac{(a+\text{d}a)^2-x^2-(a+\text{d}a)^2+y^2}{a^2-x^2-a^2+y^2}  \\ \times \dfrac{\sqrt{a^2-x^2}+\sqrt{a^2-y^2}}{\sqrt{(a+\text{d}a)^2-x^2}+\sqrt{(a+\text{d}a)^2-y^2}} \biggr|\\
    = - \log\left|\dfrac{\sqrt{(a+\text{d}a)^2-x^2}+\sqrt{(a+\text{d}a)^2-y^2}}{\sqrt{a^2-x^2}+\sqrt{a^2-y^2}} \right|.
\end{multline}

\noindent This expression can be plugged back into \eqref{Z_difference} to obtain,

\begin{multline}
    Z(\frac{x}{a+\text{d}a}, \frac{y}{a+\text{d}a}) - Z(x/a, y/a) = \\ \log\left|\dfrac{\sqrt{(a+\text{d}a)^2-x^2}+\sqrt{(a+\text{d}a)^2-y^2}}{\sqrt{(a+\text{d}a)^2-x^2}-\sqrt{(a+\text{d}a)^2-y^2}}\right| \\ - 
    \log\left|\dfrac{\sqrt{a^2-x^2}+\sqrt{a^2-y^2}}{\sqrt{a^2-x^2}-\sqrt{a^2-y^2}}\right|\\
    = 2\log\left|\dfrac{\sqrt{(a+\text{d}a)^2-x^2}+\sqrt{(a+\text{d}a)^2-y^2}}{\sqrt{a^2-x^2}+\sqrt{a^2-y^2}}\right|.
\end{multline}

\noindent Hence,

\begin{multline}
    \dfrac{1}{\text{d}a}\left(  Z(\frac{x}{a+\text{d}a}, \frac{y}{a+\text{d}a}) - Z(x/a, y/a) \right) \\
    = \dfrac{2}{\text{d}a}\log\left|\dfrac{\sqrt{(a+\text{d}a)^2-x^2}+\sqrt{(a+\text{d}a)^2-y^2}}{\sqrt{a^2-x^2}+\sqrt{a^2-y^2}}\right|. 
\end{multline}

\noindent Now, the terms in the numerator can be expanded with a Taylor series,

\begin{equation}
    \sqrt{(a+\text{d}a)^2-x^2} = \sqrt{a^2-x^2} + \dfrac{a}{\sqrt{a^2-x^2}}\text{d}a + O(\text{d}a^2),
\end{equation}

\noindent leading to,

\begin{multline}
    \dfrac{1}{\text{d}a}\left(  Z(\frac{x}{a+\text{d}a}, \frac{y}{a+\text{d}a}) - Z(x/a, y/a) \right) = \\
    \dfrac{2}{\text{d}a}\log\Biggl|\dfrac{\sqrt{a^2-x^2} + \dfrac{a}{\sqrt{a^2-x^2}}\text{d}a + O(\text{d}a^2)}{\sqrt{a^2-x^2}+\sqrt{a^2-y^2}} \\ + \dfrac{\sqrt{a^2-y^2} + \dfrac{a}{\sqrt{a^2-y^2}}\text{d}a + O(\text{d}a^2)}{\sqrt{a^2-x^2}+\sqrt{a^2-y^2}}\Biggr|,
\end{multline}

\noindent which, after using a Taylor expansion, simplifies to, 

\begin{multline}
    \dfrac{1}{\text{d}a}\left(  Z(\frac{x}{a+\text{d}a}, \frac{y}{a+\text{d}a}) - Z(x/a, y/a) \right) \\ =  \dfrac{2a}{\sqrt{a^2-x^2}\sqrt{a^2-y^2}} + O(\text{d}a).
\end{multline}

\noindent Now, we can finally go back to the energy release rate,
    
\begin{multline}
    2G = -\dfrac{dU}{\text{d}a} \\ = \dfrac{4}{\pi E'}\int_{0}^{a} p^{a}(x)\int_0^{a} p^{a}(y)\dfrac{2a}{\sqrt{a^2-x^2}\sqrt{a^2-y^2}} \text{d}y\text{d}x \\ =
    \dfrac{8a}{\pi E'}\int_{0}^{1}\dfrac{p^{a}(ar)}{\sqrt{1-r^2}}\int_0^{1} \dfrac{p^{a}(as)}{\sqrt{1-s^2}} \text{d}s\text{d}r
    \\ = \dfrac{8a}{\pi E'}\left( \int_{0}^{1}\dfrac{p^{a}(as)}{\sqrt{1-s^2}}\text{d}s\right)^2.
\end{multline}

\noindent From \cite{bazant2019fracture}, the stress intensity factor under these same conditions is,

\begin{equation}
    K_I = 2\sqrt{\dfrac{a}{\pi}}\left( \int_{0}^{1}\dfrac{p^{a}(as)}{\sqrt{1-s^2}}\text{d}s\right)
\end{equation}

\noindent From a simple inspection, one can see that $G = K_I^2/E'$, which guarantees the equivalence of the energy release rate criteria under the assumption in Figure \ref{fig:dry_crack} and the stress intensity factor condition. If instead, one assumes that the pressure load in the vicinity of a propagating crack behaves as in Figure \ref{fig:wet_crack}, this equivalence between the energetic criteria and the stress intensity factor may be violated.

\newpage

\bibliographystyle{elsarticle-num} 
\bibliography{cas-refs}

\begin{thebibliography}{10}
\expandafter\ifx\csname url\endcsname\relax
  \def\url#1{\texttt{#1}}\fi
\expandafter\ifx\csname urlprefix\endcsname\relax\def\urlprefix{URL }\fi
\expandafter\ifx\csname href\endcsname\relax
  \def\href#1#2{#2} \def\path#1{#1}\fi

\bibitem{li2015review}
Q.~Li, H.~Xing, J.~Liu, X.~Liu, A review on hydraulic fracturing of
  unconventional reservoir, Petroleum 1~(1) (2015) 8--15.

\bibitem{mair2012shale}
R.~Mair, M.~Bickle, D.~Goodman, B.~Koppelman, J.~Roberts, R.~Selley,
  Z.~Shipton, H.~Thomas, A.~Walker, E.~Woods, et~al., Shale gas extraction in
  the {UK}: a review of hydraulic fracturing, technical report (2012).

\bibitem{shinmura1997fluid}
A.~Shinmura, V.~E. Saouma, Fluid fracture interaction in pressurized reinforced
  concrete vessels, Materials and Structures 30~(2) (1997) 72--80.

\bibitem{wang2017experimental}
Y.~Wang, J.~Jia, Experimental study on the influence of hydraulic fracturing on
  high concrete gravity dams, Engineering Structures 132 (2017) 508--517.

\bibitem{capps2021critical}
N.~Capps, C.~Jensen, F.~Cappia, J.~Harp, K.~Terrani, N.~Woolstenhulme,
  D.~Wachs, A critical review of high burnup fuel fragmentation, relocation,
  and dispersal under loss-of-coolant accident conditions, Journal of Nuclear
  Materials 546 (2021) 152750.

\bibitem{turnbull2015assessment}
J.~Turnbull, S.~Yagnik, M.~Hirai, D.~Staicu, C.~Walker, An assessment of the
  fuel pulverization threshold during {LOCA}-type temperature transients,
  Nuclear Science and Engineering 179~(4) (2015) 477--485.

\bibitem{bourdin2000numerical}
B.~Bourdin, G.~A. Francfort, J.-J. Marigo, Numerical experiments in revisited
  brittle fracture, Journal of the Mechanics and Physics of Solids 48~(4)
  (2000) 797--826.

\bibitem{bourdin2012variational}
B.~Bourdin, C.~P. Chukwudozie, K.~Yoshioka, et~al., A variational approach to
  the numerical simulation of hydraulic fracturing, in: SPE Annual Technical
  Conference and Exhibition, Society of Petroleum Engineers, 2012.

\bibitem{wheeler2014augmented}
M.~F. Wheeler, T.~Wick, W.~Wollner, An augmented-lagrangian method for the
  phase-field approach for pressurized fractures, Computer Methods in Applied
  Mechanics and Engineering 271 (2014) 69--85.

\bibitem{mikelic2015quasi}
A.~Mikeli{\'c}, M.~F. Wheeler, T.~Wick, A quasi-static phase-field approach to
  pressurized fractures, Nonlinearity 28~(5) (2015) 1371.

\bibitem{peco2017influence}
C.~Peco, W.~Chen, Y.~Liu, M.~Bandi, J.~E. Dolbow, E.~Fried, Influence of
  surface tension in the surfactant-driven fracture of closely-packed
  particulate monolayers, Soft Matter 13~(35) (2017) 5832--5841.

\bibitem{jiang2022phase}
W.~Jiang, T.~Hu, L.~K. Aagesen, S.~Biswas, K.~A. Gamble, A phase-field model of
  quasi-brittle fracture for pressurized cracks: Application to {UO2}
  high-burnup microstructure fragmentation, Theoretical and Applied Fracture
  Mechanics 119 (2022) 103348.

\bibitem{hu2021variationalthesis}
T.~Hu, A variational framework for phase-field fracture modeling with
  applications to fragmentation, desiccation, ductile failure, and spallation,
  Ph.D. thesis, Duke University (2021).

\bibitem{karma2001phase}
A.~Karma, D.~A. Kessler, H.~Levine, Phase-field model of mode iii dynamic
  fracture, Physical Review Letters 87~(4) (2001) 045501.

\bibitem{francfort1998revisiting}
G.~A. Francfort, J.-J. Marigo, Revisiting brittle fracture as an energy
  minimization problem, Journal of the Mechanics and Physics of Solids 46~(8)
  (1998) 1319--1342.

\bibitem{ambrosio1990approximation}
L.~Ambrosio, V.~M. Tortorelli, Approximation of functional depending on jumps
  by elliptic functional via t-convergence, Communications on Pure and Applied
  Mathematics 43~(8) (1990) 999--1036.

\bibitem{alessi2014gradient}
R.~Alessi, J.-J. Marigo, S.~Vidoli, Gradient damage models coupled with
  plasticity and nucleation of cohesive cracks, Archive for Rational Mechanics
  and Analysis 214~(2) (2014) 575--615.

\bibitem{ambati2015phase}
M.~Ambati, T.~Gerasimov, L.~De~Lorenzis, Phase-field modeling of ductile
  fracture, Computational Mechanics 55~(5) (2015) 1017--1040.

\bibitem{miehe2016phase}
C.~Miehe, S.~Mauthe, Phase field modeling of fracture in multi-physics
  problems. part iii. crack driving forces in hydro-poro-elasticity and
  hydraulic fracturing of fluid-saturated porous media, Computer Methods in
  Applied Mechanics and Engineering 304 (2016) 619--655.

\bibitem{borden2016phase}
M.~J. Borden, T.~J. Hughes, C.~M. Landis, A.~Anvari, I.~J. Lee, A phase-field
  formulation for fracture in ductile materials: Finite deformation balance law
  derivation, plastic degradation, and stress triaxiality effects, Computer
  Methods in Applied Mechanics and Engineering 312 (2016) 130--166.

\bibitem{hu2021variationalpaper}
T.~Hu, B.~Talamini, A.~J. Stershic, M.~R. Tupek, J.~E. Dolbow, A variational
  phase-field model for ductile fracture with coalescence dissipation,
  Computational Mechanics 68~(2) (2021) 311--335.

\bibitem{wilson2016phase}
Z.~A. Wilson, C.~M. Landis, Phase-field modeling of hydraulic fracture, Journal
  of the Mechanics and Physics of Solids 96 (2016) 264--290.

\bibitem{chukwudozie2019variational}
C.~Chukwudozie, B.~Bourdin, K.~Yoshioka, A variational phase-field model for
  hydraulic fracturing in porous media, Computer Methods in Applied Mechanics
  and Engineering 347 (2019) 957--982.

\bibitem{mikelic2015phase1}
A.~Mikeli{\'c}, M.~F. Wheeler, T.~Wick, Phase-field modeling of a fluid-driven
  fracture in a poroelastic medium, Computational Geosciences 19~(6) (2015)
  1171--1195.

\bibitem{santillan2018phase}
D.~Santill{\'a}n, R.~Juanes, L.~Cueto-Felgueroso, Phase field model of
  hydraulic fracturing in poroelastic media: Fracture propagation, arrest, and
  branching under fluid injection and extraction, Journal of Geophysical
  Research: Solid Earth 123~(3) (2018) 2127--2155.

\bibitem{maurini2013crack}
C.~Maurini, B.~Bourdin, G.~Gauthier, V.~Lazarus, Crack patterns obtained by
  unidirectional drying of a colloidal suspension in a capillary tube:
  experiments and numerical simulations using a two-dimensional variational
  approach, International Journal of Fracture 184~(1-2) (2013) 75--91.

\bibitem{heider2020phase}
Y.~Heider, W.~Sun, A phase field framework for capillary-induced fracture in
  unsaturated porous media: Drying-induced vs. hydraulic cracking, Computer
  Methods in Applied Mechanics and Engineering 359 (2020) 112647.

\bibitem{cajuhi2018phase}
T.~Cajuhi, L.~Sanavia, L.~De~Lorenzis, Phase-field modeling of fracture in
  variably saturated porous media, Computational Mechanics 61~(3) (2018)
  299--318.

\bibitem{hu2020frictionless}
H.~Tianchen, J.~Guilleminot, J.~Dolbow, A phase-field model of fracture with
  frictionless contact and random fracture properties: Application to thin-film
  fracture and soil desiccation, Computer Methods in Applied Mechanics and
  Engineering 368 (2020).

\bibitem{bourdin2011time}
B.~Bourdin, C.~J. Larsen, C.~L. Richardson, A time-discrete model for dynamic
  fracture based on crack regularization, International Journal of Fracture
  168~(2) (2011) 133--143.

\bibitem{borden2012phase}
M.~J. Borden, C.~V. Verhoosel, M.~A. Scott, T.~J. Hughes, C.~M. Landis, A
  phase-field description of dynamic brittle fracture, Computer Methods in
  Applied Mechanics and Engineering 217 (2012) 77--95.

\bibitem{hofacker2013phase}
M.~Hofacker, C.~Miehe, A phase field model of dynamic fracture: Robust field
  updates for the analysis of complex crack patterns, International Journal for
  Numerical Methods in Engineering 93~(3) (2013) 276--301.

\bibitem{schluter2014phase}
A.~Schl{\"u}ter, A.~Willenb{\"u}cher, C.~Kuhn, R.~M{\"u}ller, Phase field
  approximation of dynamic brittle fracture, Computational Mechanics 54~(5)
  (2014) 1141--1161.

\bibitem{li2016gradient}
T.~Li, J.-J. Marigo, D.~Guilbaud, S.~Potapov, Gradient damage modeling of
  brittle fracture in an explicit dynamics context, International Journal for
  Numerical Methods in Engineering 108~(11) (2016) 1381--1405.

\bibitem{kamensky2018hyperbolic}
D.~Kamensky, G.~Moutsanidis, Y.~Bazilevs, Hyperbolic phase field modeling of
  brittle fracture: Part i—theory and simulations, Journal of the Mechanics
  and Physics of Solids 121 (2018) 81--98.

\bibitem{moutsanidis2018hyperbolic}
G.~Moutsanidis, D.~Kamensky, J.~Chen, Y.~Bazilevs, Hyperbolic phase field
  modeling of brittle fracture: Part ii—immersed iga--rkpm coupling for
  air-blast--structure interaction, Journal of the Mechanics and Physics of
  Solids 121 (2018) 114--132.

\bibitem{wu2020fracture}
C.~Wu, J.~Fang, Z.~Zhang, A.~Entezari, G.~Sun, M.~V. Swain, Q.~Li, Fracture
  modeling of brittle biomaterials by the phase-field method, Engineering
  Fracture Mechanics 224 (2020) 106752.

\bibitem{raina2016phase}
A.~Raina, C.~Miehe, A phase-field model for fracture in biological tissues,
  Biomechanics and modeling in mechanobiology 15~(3) (2016) 479--496.

\bibitem{nagaraja2021phase}
S.~Nagaraja, K.~Leichsenring, M.~Ambati, L.~De~Lorenzis, M.~B{\"o}l, On a
  phase-field approach to model fracture of small intestine walls, Acta
  Biomaterialia 130 (2021) 317--331.

\bibitem{gultekin2016phase}
O.~G{\"u}ltekin, H.~Dal, G.~A. Holzapfel, A phase-field approach to model
  fracture of arterial walls: theory and finite element analysis, Computer
  Methods in Applied Mechanics and Engineering 312 (2016) 542--566.

\bibitem{gultekin2018numerical}
O.~G{\"u}ltekin, H.~Dal, G.~A. Holzapfel, Numerical aspects of anisotropic
  failure in soft biological tissues favor energy-based criteria: A
  rate-dependent anisotropic crack phase-field model, Computer Methods in
  Applied Mechanics and Engineering 331 (2018) 23--52.

\bibitem{ambati2015review}
M.~Ambati, T.~Gerasimov, L.~De~Lorenzis, A review on phase-field models of
  brittle fracture and a new fast hybrid formulation, Computational Mechanics
  55~(2) (2015) 383--405.

\bibitem{wu2020phase}
J.-Y. Wu, V.~P. Nguyen, C.~T. Nguyen, D.~Sutula, S.~Sinaie, S.~P. Bordas,
  Phase-field modeling of fracture, in: Advances in Applied Mechanics, Vol.~53,
  Elsevier, 2020, pp. 1--183.

\bibitem{francfort2021variational}
G.~Francfort, Variational fracture: twenty years after, International Journal
  of Fracture (2021) 1--11.

\bibitem{tanne2022loss}
E.~Tann{\'e}, B.~Bourdin, K.~Yoshioka, On the loss of symmetry in toughness
  dominated hydraulic fractures, International Journal of Fracture 237~(1-2)
  (2022) 189--202.

\bibitem{zulian2021large}
P.~Zulian, A.~Kopani{\v{c}}{\'a}kov{\'a}, M.~G.~C. Nestola, A.~Fink, N.~A.
  Fadel, J.~VandeVondele, R.~Krause, Large scale simulation of pressure induced
  phase-field fracture propagation using utopia, CCF transactions on high
  performance computing 3~(4) (2021) 407--426.

\bibitem{yoshioka2019comparative}
K.~Yoshioka, F.~Parisio, D.~Naumov, R.~Lu, O.~Kolditz, T.~Nagel, Comparative
  verification of discrete and smeared numerical approaches for the simulation
  of hydraulic fracturing, GEM-International Journal on Geomathematics 10~(1)
  (2019) 13.

\bibitem{yoshioka2020crack}
K.~Yoshioka, D.~Naumov, O.~Kolditz, On crack opening computation in variational
  phase-field models for fracture, Computer Methods in Applied Mechanics and
  Engineering 369 (2020) 113210.

\bibitem{heider2017phase}
Y.~Heider, B.~Markert, A phase-field modeling approach of hydraulic fracture in
  saturated porous media, Mechanics Research Communications 80 (2017) 38--46.

\bibitem{li2022hydro}
H.~Li, H.~Lei, Z.~Yang, J.~Wu, X.~Zhang, S.~Li, A hydro-mechanical-damage fully
  coupled cohesive phase field model for complicated fracking simulations in
  poroelastic media, Computer Methods in Applied Mechanics and Engineering 399
  (2022) 115451.

\bibitem{heider2021review}
Y.~Heider, A review on phase-field modeling of hydraulic fracturing,
  Engineering Fracture Mechanics 253 (2021) 107881.

\bibitem{lorentz2011convergence}
E.~Lorentz, S.~Cuvilliez, K.~Kazymyrenko, Convergence of a gradient damage
  model toward a cohesive zone model, Comptes Rendus M{\'e}canique 339~(1)
  (2011) 20--26.

\bibitem{geelen2019phase}
R.~J. Geelen, Y.~Liu, T.~Hu, M.~R. Tupek, J.~E. Dolbow, A phase-field
  formulation for dynamic cohesive fracture, Computer Methods in Applied
  Mechanics and Engineering 348 (2019) 680--711.

\bibitem{wu2017unified}
J.-Y. Wu, A unified phase-field theory for the mechanics of damage and
  quasi-brittle failure, Journal of the Mechanics and Physics of Solids 103
  (2017) 72--99.

\bibitem{sicsic2013gradient}
P.~Sicsic, J.-J. Marigo, From gradient damage laws to griffith’s theory of
  crack propagation, Journal of Elasticity 113~(1) (2013) 55--74.

\bibitem{ballarini2016closed}
R.~Ballarini, G.~Royer-Carfagni, Closed-path j-integral analysis of bridged and
  phase-field cracks, Journal of Applied Mechanics 83~(6) (2016).

\bibitem{hossain2014effective}
M.~Hossain, C.-J. Hsueh, B.~Bourdin, K.~Bhattacharya, Effective toughness of
  heterogeneous media, Journal of the Mechanics and Physics of Solids 71 (2014)
  15--32.

\bibitem{li1985comparison}
F.~Z. Li, C.~F. Shih, A.~Needleman, A comparison of methods for calculating
  energy release rates, Engineering Fracture Mechanics 21~(2) (1985) 405--421.

\bibitem{shih1986energy}
C.~Shih, B.~Moran, T.~Nakamura, Energy release rate along a three-dimensional
  crack front in a thermally stressed body, International Journal of Fracture
  30~(2) (1986) 79--102.

\bibitem{amor2009regularized}
H.~Amor, J.-J. Marigo, C.~Maurini, Regularized formulation of the variational
  brittle fracture with unilateral contact: Numerical experiments, Journal of
  the Mechanics and Physics of Solids 57~(8) (2009) 1209--1229.

\bibitem{miehe2010phase}
C.~Miehe, M.~Hofacker, F.~Welschinger, A phase field model for rate-independent
  crack propagation: Robust algorithmic implementation based on operator
  splits, Computer Methods in Applied Mechanics and Engineering 199~(45-48)
  (2010) 2765--2778.

\bibitem{detournay2016mechanics}
E.~Detournay, Mechanics of hydraulic fractures, Annual Review of Fluid
  Mechanics 48 (2016) 311--339.

\bibitem{garagash2000tip}
D.~Garagash, E.~Detournay, The tip region of a fluid-driven fracture in an
  elastic medium, Journal of Applied Mechanics 67~(1) (2000) 183--192.

\bibitem{detournay2004propagation}
E.~Detournay, Propagation regimes of fluid-driven fractures in impermeable
  rocks, International Journal of Geomechanics 4~(1) (2004) 35--45.

\bibitem{garagash2005plane}
D.~I. Garagash, E.~Detournay, Plane-strain propagation of a fluid-driven
  fracture: small toughness solution, Journal of Applied Mechanics 72~(6)
  (2005) 916--928.

\bibitem{bunger2005toughness}
A.~P. Bunger, E.~Detournay, D.~I. Garagash, Toughness-dominated hydraulic
  fracture with leak-off, International Journal of Fracture 134~(2) (2005)
  175--190.

\bibitem{jiang2020three}
W.~Jiang, T.~Hu, L.~K. Aagesen, Y.~Zhang, Three-dimensional phase-field
  modeling of porosity dependent intergranular fracture in {UO2}, Computational
  Materials Science 171 (2020) 109269.

\bibitem{lorentz2011gradient}
E.~Lorentz, V.~Godard, Gradient damage models: Toward full-scale computations,
  Computer Methods in Applied Mechanics and Engineering 200~(21-22) (2011)
  1927--1944.

\bibitem{rice1968mathematical}
J.~R. Rice, et~al., Mathematical analysis in the mechanics of fracture,
  Fracture: an advanced treatise 2 (1968) 191--311.

\bibitem{rice1968path}
J.~R. Rice, A path independent integral and the approximate analysis of strain
  concentration by notches and cracks, Journal of Applied Mechanics 35~(2)
  (1968) 379--386.

\bibitem{karlsson1978jintegral}
A.~Karlsson, J.~Backlund, J-integral at loaded crack surfaces, International
  Journal of Fracture 14 (1978) 311--318.

\bibitem{karush1939minima}
W.~Karush, Minima of functions of several variables with inequalities as side
  constraints, M. Sc. Dissertation. Dept. of Mathematics, Univ. of Chicago
  (1939).

\bibitem{kuhn1951nonlinear}
H.~W. Kuhn, A.~W. Tucker, Nonlinear {P}rogramming, in: Proceedings of the
  {S}econd {B}erkeley {S}ymposium on {M}athematical {S}tatistics and
  {P}robability, University of California Press, 1951, pp. 481--492.

\bibitem{heister2015primal}
T.~Heister, M.~F. Wheeler, T.~Wick, A primal-dual active set method and
  predictor-corrector mesh adaptivity for computing fracture propagation using
  a phase-field approach, Computer Methods in Applied Mechanics and Engineering
  290 (2015) 466--495.

\bibitem{hu2020phase}
T.~Hu, J.~Guilleminot, J.~E. Dolbow, A phase-field model of fracture with
  frictionless contact and random fracture properties: Application to thin-film
  fracture and soil desiccation, Computer Methods in Applied Mechanics and
  Engineering 368 (2020) 113106.

\bibitem{raccoon}
T.~Hu, \href{https://hugary1995.github.io/raccoon/index.html}{{RACCOON} - a
  parallel finite-element code specialized in phase-field for fracture} (2020).
\newline\urlprefix\url{https://hugary1995.github.io/raccoon/index.html}

\bibitem{gaston2009moose}
D.~Gaston, C.~Newman, G.~Hansen, D.~Lebrun-Grandie, Moose: A parallel
  computational framework for coupled systems of nonlinear equations, Nuclear
  Engineering and Design 239~(10) (2009) 1768--1778.

\bibitem{permann2020moose}
C.~J. Permann, D.~R. Gaston, D.~Andr{\v{s}}, R.~W. Carlsen, F.~Kong, A.~D.
  Lindsay, J.~M. Miller, J.~W. Peterson, A.~E. Slaughter, R.~H. Stogner,
  et~al., Moose: Enabling massively parallel multiphysics simulation, SoftwareX
  11 (2020) 100430.

\bibitem{lindsay20222}
A.~D. Lindsay, D.~R. Gaston, C.~J. Permann, J.~M. Miller, D.~Andr{\v{s}}, A.~E.
  Slaughter, F.~Kong, J.~Hansel, R.~W. Carlsen, C.~Icenhour, et~al., 2.0-moose:
  Enabling massively parallel multiphysics simulation, SoftwareX 20 (2022)
  101202.

\bibitem{stoeckhert2015fracture}
F.~Stoeckhert, M.~Molenda, S.~Brenne, M.~Alber, Fracture propagation in
  sandstone and slate--laboratory experiments, acoustic emissions and fracture
  mechanics, Journal of Rock Mechanics and Geotechnical Engineering 7~(3)
  (2015) 237--249.

\bibitem{irwin1957analysis}
G.~R. Irwin, {Analysis of Stresses and Strains Near the End of a Crack
  Traversing a Plate}, Journal of Applied Mechanics 24~(3) (1957) 361--364.

\bibitem{fosdick2003}
R.~Fosdick, L.~Truskinovsky, {About Clapeyron's Theorem in Linear Elasticity},
  Journal of Elasticity 72 (2003) 145--172.

\bibitem{bazant2019fracture}
Z.~P. Bazant, J.~Planas, Fracture and size effect in concrete and other
  quasibrittle materials, Routledge, 2019.

\end{thebibliography}

\end{document}